\documentclass[aps,pre,reprint,superscriptaddress,floatfix]{revtex4-2}
\usepackage{blindtext}
\usepackage[utf8]{inputenc}
\usepackage{graphicx,dcolumn,bm,gensymb,amsmath,amssymb,amsthm,adjustbox,wasysym,upgreek}
\graphicspath{{figures/}}
\usepackage{fourier}
\usepackage[T1]{fontenc}

\usepackage[dvipsnames]{xcolor}
\usepackage{totcount}
\newcounter{tofixn}

\regtotcounter{tofixn}

\usepackage[normalem]{ulem}

\newcommand{\E}{{\rm e}}

\usepackage[inline]{enumitem}
\usepackage{xspace}
\newcommand{\ie}{\emph{i.e.}\xspace}
\newcommand{\eg}{\emph{e.g.}\xspace}

\newcommand{\apriori}{\emph{a priori}\xspace}

\newcommand{\RES}{\text{Re}_\text{s}}

\graphicspath{{./}}

\begin{document}
\title{Modeling fast acoustic streaming: steady state and transient flow solutions}
\author{Jeremy Orosco}
\author{James Friend}
\affiliation{Medically Advanced Devices Laboratory, Department of Mechanical and Aerospace Engineering, Jacobs School of Engineering, and Department of Surgery, School of Medicine, University of California San Diego, 9500 Gilman Dr.\ MC0411 La Jolla, CA 92093 USA}
\email{jfriend@ucsd.edu}
\homepage{http://friend.ucsd.edu}

\date{Received \today; published }
\begin{abstract}
    
    Traditionally, acoustic streaming is assumed to be a steady-state, relatively slow fluid response to passing acoustic waves. This assumption, the so-called \emph{slow streaming} assumption, was made over 150 years ago by Lord Rayleigh. It produces a tractable asymptotic perturbation analysis from the nonlinear governing equations, separating the acoustic field from the acoustic streaming that it generates. Unfortunately, this assumption is generally invalid in the modern microacoustofluidics context, where the fluid flow and acoustic particle velocities are comparable. Despite this issue, the assumption is still widely used today, as there is no suitable alternative.

    We describe a novel mathematical method to supplant the classic approach and properly treat the spatiotemporal scale disparities present between the acoustics and remaining fluid dynamics. The method is applied in this work to well-known problems of semi-infinite extent defined by the Navier-Stokes equations, and preserves unsteady fluid behavior driven by the acoustic wave. The separation of the governing equations between the fast (acoustic) and slow (hydrodynamic) spatiotemporal scales are shown to naturally arise from the intrinsic properties of the fluid under forcing, not by arbitrary assumption beforehand. Solution of the unsteady streaming field equations provides physical insight into observed temporal evolution of bulk streaming flows that, to date, have not been modeled. A Burgers equation is derived from the new method to represent unsteady flow. By then assuming steady flow,  a Riccati equation is found to represent it. Solving these equations produces direct, concise insight into the nonlinearity of the acoustic streaming phenomenon alongside an absolute, universal upper bound of 50\% for the energy efficiency in transducing acoustic energy input to the acoustic streaming energy output. Rigorous validation with respect to experimental and theoretical results from the classic literature is presented to connect this work to past efforts by many authors.

\end{abstract}
\maketitle
%
%
\section{Introduction}
The simple act of passing an acoustic wave through a fluid produces of a useful flow within---acoustic streaming \citep{friend_microscale_2011}---and a complex physical process responsible for it. Counterintuitive fluid behaviors often appear from acoustic streaming, behaviors that turn out to have practical applications, including biosensing~\citep{bussonniere_acoustic_2020,orazbayev_far-field_2020}, medical diagnostics~\citep{karthick_improved_2018,zhang_microliter_2021}, nozzle-free printing~\citep{connacher_droplet_2020}, smart materials~\citep{gibaud_rheoacoustic_2020}, gene editing~\citep{belling_acoustofluidic_2020}, energy storage~\citep{huang_enabling_2020,lajoinie_high-frequency_2021}, drug delivery~\citep{benmore_amorphization_2011,blamey_microscale_2013}, noise insulation~\citep{xu_physical_2020}, and a great many others~\citep{friend_microscale_2011,plaksin_intramembrane_2014,yang_topological_2015,connacher_micro/nano_2018,bach_suppression_2020}. Taken together, this broad range of utility and the phenomena responsible for it indicate the importance of deriving consistent theoretical representations for explaining acoustic streaming. Unfortunately, its analysis is not straightforward.

The analysis of acoustic streaming is difficult for several reasons. First, the fundamental conversion mechanism from an acoustic wave to acoustic streaming is represented by the nonlinear term in the Navier-Stokes equation, the streamwise acceleration or Reynolds stress. Elimination of this term is often the first step of many other solution approaches, but the term must be retained here. Second, other typical assumptions including steady, inviscid, or incompressible flow cannot generally be applied to acoustic streaming, even in the modern context where micro to nano-scale fluid phenomena are usually considered. Finally, there is a large discrepancy in the spatiotemporal scales between the acoustic field and the fluid dynamics. The acoustic field  occurs at fast and small scales, yet drives fluid dynamics at much slower and larger scales. This precludes direct numerical solution of the governing Navier-Stokes equations, since discretization sufficient to model the acoustics is computationally prohibitive.

In the past, formal asymptotic expansions have been almost exclusively used with a relevant small parameter, usually the acoustic Mach number, to decompose the mathematical representation of the flow field. A tractable set of equations can be produced from the  nonlinear partial differential equations (PDEs) that define the mass, momentum, and sometimes energy conservation. In this approach, the dependent variables---pressure, density, velocity, and sometimes temperature---are expanded in a Taylor series in terms of the small parameter. The zeroth-order terms represent the fluid dynamics not associated with the acoustics or the acoustically-driven phenomena, the first-order terms represent the acoustic field, and the second-order terms represent the nonlinear portion of the acoustics that gives rise to acoustic streaming. Higher order terms are ignored. The acoustic streaming flows are assumed to be unchanging in time, the steady result of time averaging the second-order terms in the series expansion. The flows are also assumed to be much slower than the particle velocities in the first order acoustic field \citep{lighthill_acoustic_1978}, the ``slow streaming'' approximation. To be clear, slow in this context is the observation that the acoustic streaming is slow relative to the particle velocity of the passing acoustic wave responsible for the streaming. It is possible to have rapid acoustic streaming and yet it still be slow relative to the particle velocity of the parent acoustic wave. This method of decomposing the flow field was first explored by Rayleigh during his study of ordered cells of recirculating flow that develop in Kundt's tubes~\citep{strutt_i._1884}. Over the years, many authors have used this approach in their analyses to produce solutions and describe acoustic streaming;  \citet{eckart_vortices_1948}, \citet{nyborg_acoustic_1965}, and \citet{westervelt_theory_1953} are especially well known.

By design, modern micro to nano-scale flows driven by acoustic fields often possess large acoustic intensities and sub-millimeter acoustic wavelengths~\citep{blamey_microscale_2013,reyt_experimental_2014}. In these systems, the overarching assumption of slow streaming may not be appropriate~\citep{kamakura_time_1996,dentry_frequency_2014,zhang_acoustic_2019}. Moreover, formal asymptotic expansions based on the slow streaming assumption become erroneous in these systems~\citep{reyt_experimental_2014,daru_acoustic_2017}. This has been known for a long time, with \citet{lighthill_acoustic_1978} notably addressing the matter in 1978. He noted that ``\ldots the question of whether a term can be neglected or not depends\ldots exclusively on its numerical magnitude and not on its mathematical order,'' referring to the slow streaming approximation as ``RNW streaming'' (after Rayleigh, Nyborg, and Westervelt). He continued, stating ``\ldots use of the [slow streaming] RNW equations will prove to be appropriate for acoustic sources of low power\ldots'' and that ``a milliwatt source would generate an RNW streaming flow much too large for the basic assumption of [slow streaming] RNW streaming theory to be satisfied'' for a propagating acoustic field at $1$\,MHz. He did provide a semi-empirical derivation for acoustic streaming that overcomes the problem, but a systematic approach has not yet been provided to model and predict acoustic streaming phenomena while avoiding some or all of the assumptions associated with the slow streaming approach since frequencies higher than $1\,$MHz are used. The point is that for many acoustic streaming phenomena, the assumptions routinely used in slow streaming are inappropriate.

There are other issues with the slow streaming representation. First, it is difficult to justify \apriori separation of fluid flows based upon the phenomena responsible for them. For example, pressure-driven flow would be represented at zeroth order, while the acoustics is at first order, and the flow phenomena responsible for acoustic streaming is at second order. There is no particular reason to believe this partitioning of the flow would properly model the fluid dynamics in a general way. Second, acoustic streaming is not a steady phenomenon. Instead, it is transient, developing over a finite time interval~\citep{kamakura_time_1996,chini_large-amplitude_2014,moudjed_scaling_2014}. And yet the transience of acoustic streaming is discarded when applying time average constraints~\citep{riley_steady_2001-1,vanneste_streaming_2011}.

Despite these issues, the classical approach remains popular today due to a lack of suitable alternatives~\citep{bailliet_acoustic_2001,vanneste_streaming_2011,riaud_influence_2017}. For example, \citet{vanneste_streaming_2011} studied a standard three-phase contact configuration and the physical progression from leaky surface acoustic waves to steady interior streaming. In their approach, they define the Eulerian mean field as the second-order component in a formal asymptotic expansion using the acoustic Mach number as the small parameter. The intrinsic assumption of convergence in that expansion must later be relaxed to permit inclusion of interior streaming amplitudes that scale with a potentially unbounded characteristic length. This can be readily observed by comparing eqs.\,(2.4), (2.7), and (4.15) in the referenced study. The length scale in question characterizes the interior flow. In practice, this length can be quite large, much larger than the system under study. 

Other approaches to the analysis of acoustic streaming have been proposed. \citet{zarembo_acoustic_1971} defined the concept of fast acoustic streaming to address concerns with how large amplitude acoustic waves could lead to a situation where ``The method of successive approximations [used in the slow streaming model]\ldots is inapplicable\ldots''. He chose to define partitions in the parameters not based upon the phenomena responsible, but instead in two parts: a time-averaged steady-state part and the remaining, transient part. However, Zarembo concluded his investigation without providing a method to solve the equations his approach produced. 

Instead of using the acoustic Mach number as the small parameter in the expansion, \citet{riley_steady_2001-1} used the inverse of a large Strouhal number. The quantity depends upon the time scale disparity between the driving acoustics and the resulting acoustic streaming flow. As this disparity increases, the representation by Riley's successive approximation expansion more accurately represents the system. The separation of time has likewise been considered in the differential operators present in the mass and momentum conservation equations. \citet{rudenko_theoretical_1977} presented a qualitative approach to time separation in these operators, while \citet{chini_large-amplitude_2014} quantified this separation to produce useful results for acoustic streaming.

A similar approach is employed by \citet{huang_enabling_2020} in the differential remapping of their arbitrary Lagrangian-Eulerian analysis of a slow streaming system with commensurate acoustic and hydrodynamic characteristic spatial scales. Their work borrows from an earlier study by \citet{xie_dynamics_2014} that illustrates a procedure for making explicit the order of separation between time scales in a slow streaming configuration. In their study, \citet{huang_enabling_2020} remark on the importance of providing a similar level of scrutiny to spatial scale disparities when they exist, though a detailed treatment of this case is left as a topic for further work. In each of the studies that explicitly considers the distinct change rates between the acoustics and hydrodynamics, averaging over the acoustic period has no effect on the streaming transience.

Indeed, many investigations have focused solely upon the temporal disparity present between two phenomena. In discrete systems where ordinary differential equations are sufficient, there is ample literature to describe the sometimes counter-intuitive effects of the temporal discrepancy \citep{rudenko_theoretical_1977,chini_large-amplitude_2014,moudjed_oscillating_2014,xie_dynamics_2014,nama_acoustic_2017}.

An alternative to all these methods is direct numerical simulation of the entire phenomena without simplification or approximation. Unfortunately, the presence of enormous disparities---typically between five and nine orders of magnitude~\citep{blamey_microscale_2013,dentry_frequency_2014}---between the spatiotemporal scales of the acoustic forcing and those associated with the streaming flow it generates make such an approach prohibitive. For example, the estimated computation time is several years to resolve a single period of fundamental motion while respecting the Nyquist criterion and using cutting-edge computational technologies~\footnote{For example, the lattice Boltzmann method with GPU-accelerated parallelization supported by several top-of-the-line graphics cards will require more than half a decade. See worksheet in [supplemental information].}. While supercomputing resources are more readily available than ever, and some analysis shortcuts can be taken, for example in approximating the boundary layer and presuming any free fluid surfaces remain stationary \citep{rezk2014poloidal}, the need to produce many solutions to understand and exploit the acoustic streaming phenomena remains an unmet challenge.

Likewise, in the laboratory setting, fully resolving the most elusive flow structures at the smallest, fastest scales remains difficult with currently available techniques and equipment. For the limited number of microacoustofluidic flows that \emph{can} be empirically characterized, one may only do so by operating the most advanced (and most costly) instrumentation, and often only after extending the novelty of some existing experimental methodology in a nontrivial way~\citep{blamey_microscale_2013, zhang_vibration_2020-1}. 


The purpose of this work is to provide a method that treats both the spatial disparities and the temporal disparities present between the acoustics and the resulting hydrodynamics. This method should also prove beneficial for problems in other disciplines, for example, the study of lasers and laser optics~\citep{mcintyre_all-optical_2010,oppo_long-term_2009}, transport in porous media~\citep{battiato_applicability_2011}, hydrogeologic modeling~\citep{scheibe_analysis_2015}, electrochemistry~\citep{magrini_extraction_2020}, structural mechanics~\citep{aubry_two-timescale_2010}, climate dynamics~\citep{robel_response_2018}, and rheology~\citep{pottier_high_2013,cates_reptation_1987}. The work proceeds as follows. In the next section, we describe the underlying notation and follow that with the necessary partitioning of the fluid dynamics into the slow, large scale (or ``streaming'') result and the fast, small scale acoustic excitation. This requires a set of physical constraints alongside the sets of equations to produce a tractable model. In section \ref{sec:bulk_streaming}, the partitioning and constraints are applied to a one-dimensional acoustic streaming jet---the classical quartz wind problem \citep{eckart_vortices_1948}. A nonlinear PDE is obtained which is then solved to produce an analytical expression for transient acoustic streaming. Simplification of this expression for the steady flow reveals a new fundamental limit governing the efficiency of energy conversion from the acoustic field to the resulting acoustic streaming flow. The results include a comprehensive comparison of our analysis with past results from experiments published in the literature. We conclude with a brief summarizing discussion.

\section{Theory}
We use a compact form of Leibniz' notation for convenience, flexibility, and clarity. To aid readability and brevity, the component (index) notation will be dropped for vector inputs, for example in the Euclidean spatial displacement $x_i=(x_1,x_2,x_3)$ that is written simply as $x$. To illustrate, we consider the vector-valued function $h_i(x,t)$ of the spatial input $x$ and the scalar temporal input $t$, where the vector output is given in index notation. We write the substantial derivative of this function as
%
\begin{align}\label{eq:substantial_derivative}
    D_t\,h_j(x,t) = d_t\,h_j(x,t)+u_i(x,t)\,d_{x,i}\,h_j(x,t).
\end{align}
Likewise, wherever implicitly or contextually understood, we will also drop arguments to functions. Moreover, the Einstein summation convention is used throughout. For example, in the dot product of the velocity $u_i(x,t)$ with the gradient of the function $h_i(x,t)$ in the last term of the previous equation. 

Since we will be working with two spatiotemporal scales, we must carefully define our differential expressions to consistently work within and across each scale. The symbol $d$ represents a derivative that is \emph{intra}dimensionally complete, but \emph{inter}dimensionally partial. The partial operator $\partial$ is reserved for derivatives that are intradimensionally partial.  For example, we later demonstrate the expansion of the time derivative as
%
\begin{align}
    d_t = \partial_t + \partial_{t}\tau\,\partial_{\tau},
\end{align}
producing the derivative across two distinct time scales: $t$ and $\tau$. 

We now assume the flow is superposed of, or ``partitioned'' into temporally slow (streaming, $(s)$) and fast (acoustic, $(a)$) components:
%
\begin{align}\label{eq:traditional_partition}
    \widetilde{u}_i(\widetilde{x},\widetilde{t}) = \widetilde{u}^{(s)}_i(\widetilde{x},\widetilde{t}) + \widetilde{u}^{(a)}_i(\widetilde{x},\widetilde{t}),
\end{align}
where we have again used index notation and where we denote all dimensional field variables, operators, and independent parameters with a tilde.

In fact, we consider a system where the magnitude of the velocity of an acoustic streaming-driven flow is on the order of the acoustically-driven particle velocity magnitude. The surface acoustic wave (SAW)-driven jet streaming described by \citet{dentry_frequency_2014,dentry_erratum_2016} is an extension of Lighthill's well-known turbulent jet model~\citep{lighthill_acoustic_1978} for use in microacoustofluidic systems. In Dentry's study and in Lighthill's study before it, the maximum streaming jet velocity, $U_s$, may generally be of the same order of magnitude as the acoustic source's particle velocity, $U_a$.

If we define the characteristic jet streaming length as $x_s$ and the corresponding streaming time scale as $t_s=x_s/U_s$, then we are free to write
%
\begin{align}
    \widetilde{u}_i &= \frac{x_s}{t_s}\,u^{(s)}_i + \xi_p\,\omega\,u^{(a)}_i,
    \label{eq:charfact}
\end{align}
where the particle displacement is $\xi_p$, and the acoustic time scale, $1/\omega$, is given in terms of the angular acoustic frequency $\omega = 2\,\pi\,f$. The absence of a tilde indicates nondimensional field variables, operators, and independent parameters. Thus, $u_i^{(s)}$ and $u_i^{(a)}$ are both $\mathcal{O}[1]$ quantities, and $x_s$, $t_s$, $\xi_p$, and $\omega$ define the relative velocity magnitudes. An important aspect of eqn.\,\eqref{eq:charfact} is the appearance of a spatial scale separation in addition to a separation in time scales. 

The nondimensionalized velocity may then be written as
%
\begin{align}\label{eq:super_nondim}
    \begin{split}
        u_i(x,\xi,t,\tau) = u^{(s)}_i(x,t) + q_p\,S\,u^{(a)}_i(x,\xi,t,\tau),
    \end{split}
\end{align}
with $q_p=\xi_p/x_s\ll1$ and $S=\omega\,t_s\gg1$, and where the nondimensional time variables $t$ and $\tau$, and the nondimensional space variables $x_i$ and $\xi_i$, are detailed further on. For now it is sufficient to note that $t$ and $x$ refer to large streaming scales, while $\tau$ and $\xi$ refer to small acoustic scales. Thus, the assumption made in eqn.\,\eqref{eq:super_nondim} is that the nondimensional streaming velocity changes only over large space scales and over long times relative to the small acoustic space and time scales. This assumption will later be shown to be valid with a comparison to experimental results, and is consistent with the acoustic streaming jets described by \citet{dentry_frequency_2014,dentry_erratum_2016} and \citet{lighthill_acoustic_1978}.
                
Since, by our earlier definition, the streaming and particle velocity magnitudes are of the same order, eqn.\,\eqref{eq:charfact} tells us that
    %
    \begin{align}\label{eq:time_space_scale_equiv}
        \frac{x_s}{\xi_p}\sim\omega\,t_s,
    \end{align}
so that $q_p\,S\sim1$. Compare this result with the outcome of assuming slow streaming, where $U_s \ll U_a$, for which $q_p\,S\gg1$.

Here, the simple and fundamental result $q_p\,S\sim1$ may be interpreted as an axiom of fast streaming upon which the remainder of the scales are developed. It implies the balance in eqn.\,\eqref{eq:time_space_scale_equiv} \emph{must follow} under fast streaming conditions.

\subsubsection{Temporal derivative partitioning}
Because a single temporal scale is insufficient to represent the acoustic streaming, we define a temporal derivative $d_t$ with dimensional form $\widetilde{d}_t$ that is complete on each time scale (fast and slow) yet is a partial derivative, splitting a time derivative between the two scales:
\begin{align}
    \widetilde{d}_t\widetilde{u}_i = \widetilde{d}_t\widetilde{u}^{(s)}_i + \widetilde{d}_t\widetilde{u}^{(a)}_i.
\end{align}
The operator $\widetilde{d}_t$ is intradimensionally complete yet interdimensionally partial. Because the fast $(\cdot)^{(a)}$ and slow $(\cdot)^{(s)}$ flow components change at drastically different rates, we define two different non-dimensional time scales in terms of the physical (``real'') time $\widetilde{t}$:
%
\begin{subequations}
    \begin{align}
    t &= \frac{1}{t_s}\,\widetilde{t},\quad \text{and}\\
    \tau &= \omega\,\widetilde{t}.
    \end{align}
\end{subequations}
Notice that $\widetilde{t} = t t_s = \tau\omega^{-1}$, such that any small change in the slow time scale produces a large change in the fast time scale. If we define $\tau = St$, $\frac{d\tau}{dt} = d_t \tau = S$ in the limit.

Differentiating a function $\chi(t,\tau)$ that is a function of the fast and slow times $\tau$ and $t$ with respect to time in the non-dimensional space requires the total differential $\widetilde{d}_t$,
%
\begin{align}\label{eq:dimensional_time_derivative}
    \begin{split}
    \widetilde{d}_t \chi&=\widetilde{\partial}_tt\,\partial_t \chi + \widetilde{\partial}_t\tau\,\partial_{\tau} \chi \\
    &=\frac{1}{t_s}\,\partial_t \chi+\omega\,\partial_{\tau} \chi.
    \end{split}
\end{align}
Likewise, the nondimensional total derivative of $\chi(t,\tau)$ is
%
\begin{align}\label{eq:primary_nondim_time_derivative}
    \begin{split}
        d_t \chi= S^{-1}\,\partial_t \chi+\partial_{\tau} \chi,
    \end{split}
\end{align}
or
%
\begin{align}\label{eq:secondary_nondim_time_derivative}
    \begin{split}
        d_t \chi= \partial_t \chi+S\,\partial_{\tau} \chi,
    \end{split}
\end{align}
depending on whether one divides eqn.\,\eqref{eq:dimensional_time_derivative} by $\omega$ or multiplies it by $t_s$, respectively.

\subsubsection{Spatial derivative partitioning}
The effects of partitioning fast and slow phenomena extend to spatial derivatives. This is dealt with in a manner similar to the temporal case, writing the dimensional gradient operator as $\widetilde{d}_{x,i}(\cdot)$. Using the partition of the flow velocity $\widetilde{u}_j$ defined in eqn.\,\eqref{eq:traditional_partition} for example, the spatial gradient of the flow velocity $\widetilde{d}_{x,i}\,\widetilde{u}_j(\widetilde{x},\widetilde{t})$ is
\begin{align}
    \widetilde{d}_{x,i}\,\widetilde{u}_j(\widetilde{x},\widetilde{t}) = \widetilde{d}_{x,i}\,\widetilde{u}^{(s)}_j(\widetilde{x},\widetilde{t}) + \widetilde{d}_{x,i}\,\widetilde{u}^{(a)}_j(\widetilde{x},\widetilde{t}).\label{eqn:dim_vel_gradient}
\end{align}
The acoustic (fast) component varies over a characteristic distance given by the wavelength $\lambda = (2\pi)(k)^{-1}$, where $k$ is the acoustic wavenumber. The slow component varies over some other length scale; in an acoustically driven jet, for example, this scale is the jet length $x_s\ll\lambda$. The observation that $x_s\ll\lambda$ is typical of bulk acoustic streaming. It may be possible to have different length scales for the fast and slow components such that the partitioning will have to be defined with a dependence upon the direction under consideration, resulting in a Hadamard product for the spatial derivative definition. In this first demonstration of this theory, we avoid the complexity by assuming one and only one spatial partition in the gradient. This is appropriate for the one-dimensional jet streaming example we provide later on, and probably would suit more general use if the ratios of the spatial scales of the acoustics to the resulting hydrodynamics do not significantly vary with respect to direction.

We define the two non-dimensional scales as 
\begin{equation}
        x_i = \frac{1}{x_s}\widetilde{x}_i\quad\text{and}\quad \xi_i = k\,\widetilde{x}_i.
\end{equation}
This helps us produce the non-dimensional gradient operator $d_{x,i}$ from eqn.\,\eqref{eqn:dim_vel_gradient}:
%
\begin{align}\label{eq:primary_nondim_spatial_derivative}
	\begin{split}
  		d_{x,i} = q_{\lambda}\,\partial_{x,i}+\partial_{\xi,i},
	\end{split}
\end{align}
where $q_{\lambda}=(k\,x_s)^{-1}\ll1$ characterizes the disparity in the fast $(a)$ and slow $(s)$ spatial scales for the one-dimensional model.

\subsubsection{Velocity partition}
We next consider a dimensional function of the dimensional space and time variables, $\widetilde{r}_i(\widetilde{x},\widetilde{t})$, and seek to partition it. Here we use, as an example, $\widetilde{r}_i(\widetilde{x},\widetilde{t})$ to mean the position of a fluid parcel in space and time and write
\begin{align}
    \begin{split}
    \widetilde{r}_i(\widetilde{x},\widetilde{t}) &= \widetilde{r}^{(s)}_i(\widetilde{x},\widetilde{t}) + \widetilde{r}^{(a)}_i(\widetilde{x},\widetilde{t}), \\
    &= x_s\,r^{(s)}_i(x,t) + \xi_p\,r^{(a)}_i(x,\xi,t,\tau).
    \end{split}
\end{align}
The nondimensional position of the same parcel may be written then as
%
\begin{align}\label{eq:nondim_trajectory_decomp}
    \begin{split}
        r_i =r^{(s)}_i+q_p\,r^{(a)}_i.
    \end{split}
\end{align}
We next use this to determine the fluid velocity through a time derivative of $r_i(x,t)$, by applying eqns.\,\eqref{eq:secondary_nondim_time_derivative}~to~\eqref{eq:nondim_trajectory_decomp}:
%
\begin{align}
    \begin{split}
        u_i = d_t\,r_i &= \partial_t\,r^{(s)}_i+ q_p\,S\,\partial_{\tau}\,r^{(a)}_i+S\,\partial_{\tau}r^{(s)}_i+ q_p\,\partial_t\,r^{(a)}_i,\\
            &\approx \partial_t\,r^{(s)}_i+ q_p\,S\,\partial_{\tau}\,r^{(a)}_i+q_p\,\partial_t\,r^{(a)}_i,
    \end{split}
\end{align}
since $\partial_{\tau}r^{(s)}_i\approx0$ (\ie, streaming motions are essentially constant with respect to acoustic time scales). Now we may write $u^{(s)}_i = \partial_t\,r^{(s)}_i$ and $u^{(a)}_i = \partial_{\tau}\,r^{(a)}_i$, leaving us with a third term in the expression
%
\begin{align}
    u_i \approx u^{(s)}_i + q_p\,S\,u^{(a)}_i + q_p\partial_{t}\,r^{(a)}_i.
    \label{eqn:nondimvelfromtraj}
\end{align}
The first two terms on the right-hand side are $\mathcal{O}\left[1\right]$ and the last term on the right-hand side is $\mathcal{O}\left[\varepsilon\right]$.

\subsubsection{Physical constraints in the nondimensional space}\label{subsec:constraints}
A complication in using the partitioning is the need to produce constraints on the partial differential equations that conserve mass and momenta at both the slow and fast scales. Because the variables in the system like the flow velocity, $u_i$, now split into two scales, $u_i^{(s)} + q_p S u_i^{(a)}$, we seek a connection between these scales that, when enforced, produces a consistent set of solutions across the two scales. 

We employ the intermediate nondimensional time $\tau_\infty$ such that $\omega^{-1}\ll\tau_\infty\ll t_s$. We assume the separation in scales is sufficient to define $\tau_\infty$ as the long time limit of the acoustic field while still leaving the slow time scale $t_s$ small enough that the transient hydrodynamics that arise from the acoustic streaming may still be preserved. 

Begin by defining the temporal average of the acoustic (fast) phenomena as
%
\begin{align}
    \begin{split}
        \left\langle\,\cdot\,\right\rangle_{\tau} &= \frac{\omega}{2\,\pi}\int^{2\,\pi/\omega}_0\lim_{\widetilde{t}\rightarrow\left(\frac{\tau_{\infty}}{\omega}\right)^{-}}(\,\cdot\,)\,d\widetilde{t} = \frac{1}{2\,\pi}\int^{2\,\pi}_0\lim_{\tau\rightarrow\tau_{\infty}^{-}}(\,\cdot\,)\,d\tau,\label{eqn:acoustic_time_average}
    \end{split}
\end{align}
and the analogous acoustic spatial average as
%
\begin{align}
    \begin{split}
        \left\langle\,\cdot\,\right\rangle_{\xi} = \left(\frac{1}{2\,\pi}\right)^3\iiint\limits^{\xi+\pi}_{\xi-\pi}(\,\cdot\,)\,d\xi'_{i,j,k}.\label{eqn:acoustic_space_average}
    \end{split}
\end{align}
The integral in eqn.\,\eqref{eqn:acoustic_space_average} is taken over a small cube of edge length $\lambda$. It represents a unity-weighted convolution that retains its dependence upon the slow (large) length scale $x$. 

Since $u_i \approx u_i^{(s)} + q_p S u_i^{(a)}$, applying the temporal average in eqn.\,\eqref{eqn:acoustic_time_average} produces $\left\langle u_i \right\rangle_\tau \approx \left\langle u_i^{(s)} \right\rangle_\tau + q_p S \left\langle u_i^{(a)} \right\rangle_\tau$. Now $\left\langle u_i^{(a)} \right\rangle_\tau = 0$ as the acoustic flow velocity will average to zero over an acoustic period $\tau$. So $\left\langle u_i \right\rangle_\tau \approx \left\langle u_i^{(s)} \right\rangle_\tau$. Similarly, we assume the acoustic wave amplitude changes slowly over a given wavelength $\lambda$, such that $\left\langle u_i^{(a)} \right\rangle_\xi = 0$. This gives 
\begin{equation}
\left\langle u_i \right\rangle_\xi \approx \left\langle u_i^{(s)} \right\rangle_\xi + q_p S \left\langle u_i^{(a)} \right\rangle_\xi = \left\langle u_i^{(s)} \right\rangle_\xi \label{eqn:spatialaverage}
\end{equation}
for the spatial average of $u_i$, equivalent to assuming the acoustic wave is periodic over the length scale $\lambda$. This condition might be violated if the damping is extreme, such that the acoustic wave amplitude significantly varies over a single wavelength. However, such a situation is rare in acoustofluidics.

Moreover, consider the effect of the temporal and spatial derivatives for the acoustic (fast) phenomena. The fast temporal derivative of $u_i$ is 
\begin{equation}
	\partial_{\tau}\,u_i \approx \partial_{\tau}\,u_i^{(s)}+q_p\,S\,\partial_{\tau}\,u^{(a)}_i.
\end{equation}
Notice that $\,u_i^{(s)} = \,u_i^{(s)}(x,t)$ so $\partial_{\tau}\,u_i^{(s)} = 0$. This gives $\partial_{\tau}\,u_i \approx q_p\,S\,\partial_{\tau}\,u^{(a)}_i$. Likewise,
\begin{equation}
	\partial_{\xi,i}\,u_j \approx \partial_{\xi,i}\,u^{(s)}_j + q_p\,S\,\partial_{\xi,i}\,u^{(a)}_j
\end{equation}
and, again, $u^{(s)}_j = u^{(s)}_j(x,t)$ alone, so $\partial_{\xi,i}\,u^{(s)}_j =0$, giving $\partial_{\xi,i}\,u_j \approx q_p\,S\,\partial_{\xi,i}\,u^{(a)}_j$. 

For example, consider acoustic jet streaming~\citep{dentry_frequency_2014}. We can model an attenuating acoustic plane wave propagating into the fluid medium aligned with the jet axis $\widetilde{x}_i$ as
%
\begin{align}\label{eq:acoustic_wave_model}
    \widetilde{u}^{(a)}(\widetilde{x},\widetilde{t}) = U_a\,\exp{[\iota\,(\kappa\,\widetilde{x}_i-\omega\,\widetilde{t})]},
\end{align}
where $\iota=\sqrt{-1}$ and the complex wavenumber is $\kappa=k+\iota\,\alpha$, given in terms of the attenuation coefficient $\alpha=\delta_a^{-1}$, with $\delta_a$ being the attenuation length. In using complex exponentials, it is assumed throughout that only the real value is retained. The magnitude of the effect of the asymmetries will depend on the extent of the attenuation over a given spatial period [see supplementary materials, Fig.\,1], so that the ratio of interest is $\kappa_i=(k\,\delta_a)^{-1}$. As $\kappa_i\rightarrow0$, the wave is negligibly attenuated over a single spatial period, so that the spatial average is zero, matching the assumption made to produce eqn.\,\eqref{eqn:spatialaverage}. We write the nondimensionalized spatial average taken in the direction of propagation as
%
\begin{align}\label{eq:0_grad_spatial_ave}
    \max_{\widetilde{x}_i\geq\pi/k}|\langle u^{(a)}_i \rangle_{\xi,i}| &= \max_{\widetilde{x}_i\geq\pi/k}\biggr|\frac{k}{2\pi}\int^{\widetilde{x}_i+\pi/k}_{\widetilde{x}_i-\pi/k}\frac{\widetilde{u}^{(a)}}{U_a}\,d\widetilde{x}_i\biggr|, \notag\\
    &= \tfrac{1}{\pi}\psi_0(\xi_{0,\mbox{\scriptsize max}})\sinh\tfrac{\pi\,\kappa_i}{\kappa_r} \approx \theta,
\end{align}
where $\psi_0(\xi_{0,\mbox{\scriptsize max}})\approx1$, and $\kappa_i\approx\theta\equiv \mu_\text{l}\,\omega/2\,\rho_0\,c^2$ when $\kappa_i\ll1$ (equivalently, when $\kappa_r\approx 1$). The nondimensional undisturbed density of the fluid is $\rho_0$. Because they will be needed later for the conservation equations, we also consider averages of up to second-order gradient fields:
%
\begin{subequations}
    \begin{align}
        \max_{\widetilde{x}_i\geq\pi/k}|\langle \partial_{\xi,i}\,u^{(a)}_i \rangle_{\xi,i}| &= \tfrac{1}{\pi}\psi_1(\xi_{1,\mbox{\scriptsize max}})\sinh\tfrac{\pi\,\kappa_i}{\kappa_r}\approx\theta,\label{eq:1_grad_spatial_ave}\\
        \max_{\widetilde{x}_i\geq\pi/k}|\langle \partial^2_{\xi,i}\,u^{(a)}_i \rangle_{\xi,i}| &= \tfrac{1}{\pi}\psi_2(\xi_{2,\mbox{\scriptsize max}})\sinh\tfrac{\pi\,\kappa_i}{\kappa_r}\approx\theta,\label{eq:2_grad_spatial_ave}
    \end{align}
\end{subequations}
which have been normalized by $U_a\,k$ and $U_a\,k^2$, respectively. When $\kappa_i\gtrsim1$, we must use the full expressions
%
\begin{subequations}
    \begin{align}
        \psi_0(\xi) &= \biggr|\frac{\E^{-\kappa_i\,\xi}(\kappa_r\sin\kappa_r\,\xi-\kappa_i\cos\kappa_r\,\xi)}{\kappa_i^2+\kappa_r^2}\biggr|,\\
        \psi_1(\xi) &=|\E^{-\kappa_i\,\xi}\cos\kappa_r\,\xi|,\\
        \psi_2(\xi) &=|\E^{-\kappa_i\,\xi}(\kappa_i\cos\kappa_r\,\xi+\kappa_r\sin\kappa_r\,\xi)|,
    \end{align}
\end{subequations}
with $\xi_{0,\mbox{\scriptsize max}}=3\,\pi/2\,\kappa_r$, $\xi_{1,\mbox{\scriptsize max}}=\pi/\kappa_r$, and $\xi_{2,\mbox{\scriptsize max}}=\pi+\arccos(\frac{2\,\kappa_i\,\kappa_r}{\kappa_i^2+\kappa_r^2})$.

\begin{figure}
    \begin{center}
        \includegraphics[width=\columnwidth]{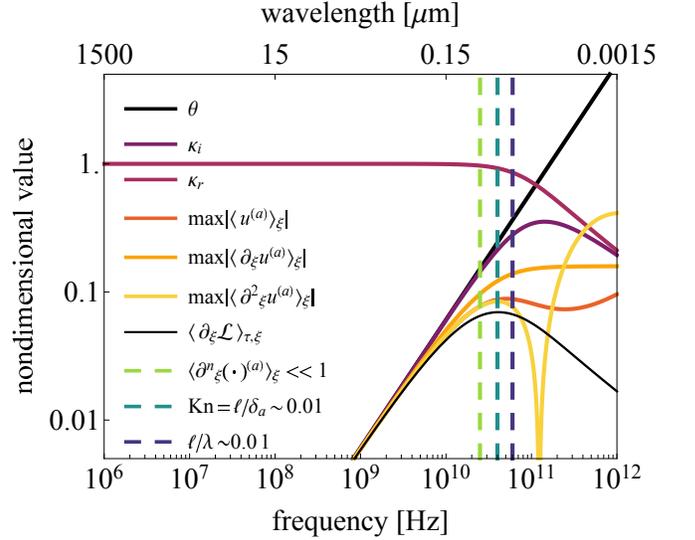}
        \caption{Important frequency-dependent nondimensional values for water (some of which are defined further on). The $\lambda\lesssim\delta_a$ limit corresponds to an upper bound on frequency of roughly $25\,$GHz (vertical, dashed, light green). Trends below this limit continue to lower frequencies unabated. Averages of the acoustic wave remain approximately valid throughout the applicable domain of continuum mechanics. The upper bound of roughly $40\,$GHz for this domain is determined by the Knudsen number $\text{Kn}\,\sim0.01$  (vertical, dashed, green), defined as the ratio of the mean free path ($\ell\approx270\,$pm) of the individual water molecules to the acoustic wave attenuation length. When $\text{Kn}\,\gtrsim0.01$, a phonon-based recasting of the representative equations may prove useful.}
        \label{fig:f01_system_scales}
    \end{center}
\end{figure}

We can now place bounds on the validity of spatial averaging by using these results. For $\kappa_i\lesssim0.15$, the normalized spatial averages in eqns.\,(\ref{eq:0_grad_spatial_ave}--\ref{eq:2_grad_spatial_ave}) are less than about 0.1. Since eqn.\,\eqref{eq:acoustic_wave_model} is a function only of the $\widetilde{x}_i$ coordinate, averaging across the remaining dimensions leaves this result unchanged. Then $\left\langle u_i \right\rangle_\xi \approx \left\langle u_i^{(s)} \right\rangle_\xi$ is valid, and as long as $\lambda\lesssim\delta_a$, $\langle \partial_{\xi,i}\,u^{(a)}_i \rangle_{\xi} \approx 0$ and $\langle \partial^2_{\xi,i}\,u^{(a)}_i \rangle_{\xi} \approx 0$. The condition $\lambda\lesssim\delta_a$ is satisfied by an acoustofluidic system operating at less than roughly $25\,$GHz (for water), as shown in Fig.\,\ref{fig:f01_system_scales}.

Virtually all acoustofluidics phenomena to date occur at $<2\,$GHz, well within the conditional range of our analysis. From an order of magnitude perspective, the valid range of the analysis exhausts the valid domain of continuum mechanics. We identify the valid range of continuum mechanics for acoustofluidics with an upper bound at $40\,$GHz. This corresponds to $\text{Kn}\sim0.01$, where we define the Knudsen number, $\text{Kn}=\ell/\delta_a$, in terms of the mean free path $\ell$. We have used the attenuation length as a characteristic scale rather than using wavelength, since this is the shorter of the two. The frequency limit defined in terms of the attenuation length is $f\sim40\,$GHz, whereas the limit defined in terms of the wavelength is $f\sim60\,$GHz.

\section{One-dimensional bulk streaming}
\label{sec:bulk_streaming}

We now apply the partitioning approach to a specific system: acoustic streaming generated by the passage of an acoustic wave into a fluid bulk from a small source. For simplicity's sake, we assume the system is one-dimensional. The dimensional, isentropic, compressible, and unsteady conservation of mass and momentum equations along the $x$-axis direction are
%
\begin{subequations}\label{eq:ns_equations}
\begin{align}
    \widetilde{d}_t\widetilde{u}_x+\widetilde{d}_x(\widetilde{\rho}\,\widetilde{u}_x)&=0,\label{eq:ns_equations_b} \\
    \widetilde{\rho}\widetilde{D}_t\widetilde{u}_x&=-\widetilde{d}_x\widetilde{P}+\mu_\text{l}\,\widetilde{d}_x^2\,\widetilde{u}_x+\widetilde{F}_x,\label{eq:ns_equations_a} 
\end{align}
\end{subequations}
with the equation of state $\widetilde{P} = \widetilde{P}(\widetilde{\rho})$ and longitudinal viscosity $\mu_\text{l} = \frac{4}{3} \mu_\text{s} + \mu_\text{V}$ written in terms of the shear viscosity, $\mu_\text{s}$, and volume viscosity, $\mu_\text{V}$ 
with longitudinal viscosity $\mu_\text{l}=\mu_\text{s}(4/3+\mu_\text{V}/\mu_\text{s})$ written in terms of the shear viscosity, $\mu_\text{s}$, and the volume viscosity, $\mu_\text{V}$. Expanding the equation of state about its nominal hydrostatic condition leads to the well-known expression
%
\begin{align}\label{eq:eq_of_state_expansion}
    \frac{\widetilde{P}'}{A}=s+\frac{B}{2A}s^2+\mathcal{O}\left[s^3\right],
\end{align}
where $\widetilde{P}'=\widetilde{P}-P_0$ and the condensation, $s=\widetilde{\rho}'/\widetilde{\rho_0}$, is given in terms of the density variation $\widetilde{\rho}'=\widetilde{\rho}-\widetilde{\rho_0}$ where $\widetilde{\rho_0}$ is the dimensional undisturbed fluid density. The term $B/A$ is Beyer's parameter characterizing the nonlinear compressibility of the fluid~\citep{beyer_nonlinear_1997}. For acoustic streaming through water, the first two terms on the right-hand side of eqn.\,\eqref{eq:eq_of_state_expansion} are at most order $\mathcal{O}\left[10^{-4}\right]$ and $\mathcal{O}\left[10^{-8}\right]$, respectively.

We continue by following \citet{riley_steady_2001-1}, differentiating eqn.\,\eqref{eq:ns_equations_a} with respect to time, substituting eqn.\,\eqref{eq:ns_equations_b} and the linear expansion of eqn.\,\eqref{eq:eq_of_state_expansion} into the result, arriving at
%
\begin{align}\label{eq:tdiff_combined_ns_equations}
&\widetilde{\rho}\,\widetilde{d}_t\widetilde{D}_t\widetilde{u}_x-\widetilde{d}_x(\widetilde{\rho}\,\widetilde{u}_x)\widetilde{D}_t\widetilde{u}_x \notag\\
&\hspace{1cm}=c^2\widetilde{d}_x^2(\widetilde{\rho}\,\widetilde{u}_x)+\mu_\text{l}\,\widetilde{d}_x^2\widetilde{d}_t\widetilde{u}_x+\widetilde{d}_t\widetilde{F}_x,
\end{align}
after eliminating the pressure from the expression as discussed above. Here $c$ is the speed of sound in the medium.

The fluid velocity $\widetilde{u}_x$ and its derivatives are present, and conveniently the derivations of the partitioning of these expressions have been provided in the previous section. However, the density, $\widetilde{\rho}$, is also present. We next consider how to partition it.

\subsection{Density partition}
We require a physically and theoretically consistent partition of the density, 
 %
 \begin{align}
	 \widetilde{\rho}(x,\xi,t,\tau) - \rho_0 = \widetilde{\rho}^{(s)}(x,t) + \widetilde{\rho}^{(a)}(x,\xi,t,\tau),\label{eqn:densitypartition}
 \end{align}
where $\rho_0$ is the unperturbed fluid density, and $\widetilde{\rho}^{(a)}$ and $\widetilde{\rho}^{(s)}$ represent density fluctuations associated with the acoustic wave and streaming-driven flow, respectively. Rather than assume in advance the nature of the partition between the acoustically and hydrodynamically-driven  density fluctuations, as is done in most models of acoustic streaming~\citep{zarembo_acoustic_1971}, we recall a fundamental result from linear acoustics \citep{beyer_nonlinear_1997,shutilov_fundamental_1988},
%
\begin{align}
	M_a\approx\max|s|,
\end{align}
valid for acoustic Mach numbers $M_a = U_a c^{-1} \ll 1$, where $U_a$ is the on-source (\ie, maximum) particle velocity of the acoustic wave. We define the magnitude of the streaming flow velocity as $U_s$ so that we may write $M_s = U_s c^{-1}$ in an analogous fashion. This motivates us to define the acoustic and streaming components to the condensation as $M_a\,\rho^{(a)}$ and $M_s\,\rho^{(s)}$, respectively, so that these terms are expected to be $\mathcal{O}[1]$. Then from $\widetilde{\rho} - \rho_0 = \widetilde{\rho}^{(s)} + \widetilde{\rho}^{(a)}$, it is natural to define the density partition
%
\begin{align}\label{eq:dimensional_density_partition}
	\widetilde{\rho}-\rho_0 =M_a\,\rho_0\,\rho^{(a)}+M_s\,\rho_0\,\rho^{(s)}.
\end{align}
The nondimensionalized density partition is then obtained by dividing eqn.\,\eqref{eq:dimensional_density_partition} by $\rho_0$ to produce the $\mathcal{O}[1]$ expression $\rho -1 = M_a\,\rho^{(a)} + M_s\,\rho^{(s)}$. Since $M_a/M_s=q_p\,S\sim 1$, the two Mach numbers are of the same order---a restatement of the order of magnitude equivalence of the particle and streaming velocities. From this result and eqn.\,\eqref{eq:dimensional_density_partition}, it can be seen that the appropriate partitioning of the density fluctuation produces an acoustic density change and a hydrodynamic density change that are also of the same order: $\widetilde{\rho}^{(s)}\sim\widetilde{\rho}^{(a)}$.

\subsection{Nondimensionalization}
We next nondimensionalize the unpartitioned conservation equations. Non-dimensionalization of eqn.\,\eqref{eq:tdiff_combined_ns_equations} produces
%
\begin{align}\label{eq:nondim_tdiff_ns_equations}
	&q_{\lambda}\,\overline{\rho}\,d_t D_t u_x-d_x(\overline{\rho}\,u_x)D_t u_x\notag\\
	&\hspace{1cm}=q_{\lambda}^2d_x^2(\overline{\rho}\,u_x)+2\,q_p\,q_{\lambda}^2\,\theta\,d_x^2d_t u_x+q_p^2\, q_{\lambda}^2\,d_tF_x,
\end{align}
where we have set $\widetilde{F}_x=\rho_0\,U_a\,k\,F_x$. We also have defined 
\begin{equation}\label{eq:modified_density_partition}
\overline{\rho} = q_p + \frac{\varepsilon\,q_p}{q_{\lambda}}\rho^{(s)} + \frac{q_p^2}{q_{\lambda}}\rho^{(a)}
\end{equation}
with advance knowledge that this auxiliary definition will simplify the upcoming partition of eqn.\,\eqref{eq:nondim_tdiff_ns_equations}. It is also helpful to combine eqns.\,\eqref{eq:ns_equations} and then nondimensionalize the result to produce
%
\begin{align}\label{eq:nondim_ns_equations}
	\overline{\rho}\,D_t u_x=-q_{\lambda}^2 d_x\overline{\rho}'+2\,q_{\lambda}^2\,q_p\,\theta\,d_x^2 u_x+q_p^2\,q_{\lambda}F_x,
\end{align}
where $\overline{\rho}'=\overline{\rho}-q_p$. We use both forms of the governing equation, eqns.\,\eqref{eq:nondim_tdiff_ns_equations}~and~\eqref{eq:nondim_ns_equations}, in the section that follows.

\subsection{Partitioning}

The traditonal small parameter expansion used in slow streaming models \citep{nyborg_acoustic_1953,vanneste_streaming_2011} is an infinite series expansion truncated beyond the second order. By contrast, substituting the partitioning expressions
$u_x\left(x,\xi,t,\tau\right) = u_x^{(s)}(x,t) + u_x^{(a)}(x,\xi,t,\tau)$ and $\rho\left(x,\xi,t,\tau\right) = \rho^{(s)}(x,t) + \rho^{(a)}(x,\xi,t,\tau)$ into eqn.\,\eqref{eq:nondim_tdiff_ns_equations} produces a finite equation without truncation. However, the complete equation produced from the partitioning for a one-dimensional acoustic streaming jet has $266$ terms. The process of partitioning that produces this equation and the complete results are provided in the Supplementary Information in a Mathematica notebook for ease of use by the reader. We regroup the terms of the equation based upon their relative scales such that
		%
		\begin{align}\label{eq:scale_ordering}
			\mathcal{O}\left[S^{-1}\right] = \mathcal{O}\left[q_p\right] \ll \mathcal{O}\left[q_{\lambda}\right] \ll 1,
		\end{align}
where $S^{-1}$ and $q_p$ are sufficiently small in comparison to $q_{\lambda}$ that, when taking successive approximations, we expand first in combinations of $S^{-1}$ and $q_p$, and subsequently in $q_{\lambda}$. The result is\\ 
            \begin{widetext}
            \resizebox{0.7\linewidth}{!}{
              \begin{minipage}{\linewidth}
                \begin{multline}\label{eq:266term}
                q_p^2 q_{\lambda }^3 \left(-2 \theta  \frac{\partial ^3u_{a,x}}{\partial \xi ^2\, \partial \tau }-\frac{\partial ^2u_{a,x}}{\partial \xi ^2}+\frac{\partial ^2u_{a,x}}{\partial \tau ^2}-\frac{\partial F_x}{\partial \tau }\right)+q_p^2 q_{\lambda }^4 \left(-4 \theta  \frac{\partial ^3u_{a,x}}{\partial x\, \partial \xi \, \partial \tau }-2 \frac{\partial ^2u_{a,x}}{\partial x\, \partial \xi }\right)+q_p^2 q_{\lambda }^5 \left(-2 \theta  \frac{\partial ^3u_{a,x}}{\partial x^2\, \partial \tau }-\frac{\partial ^2u_{a,x}}{\partial x^2}\right)+\varepsilon  q_p q_{\lambda }^3 \left(-2 \theta  \frac{\partial ^3u_{s,x}}{\partial \xi ^2\, \partial \tau }-\frac{\partial ^2u_{s,x}}{\partial \xi ^2}+\frac{\partial ^2u_{s,x}}{\partial \tau ^2}\right)+\\
                \varepsilon  q_p q_{\lambda }^4 \left(-4 \theta  \frac{\partial ^3u_{s,x}}{\partial x\, \partial \xi \, \partial \tau }-2 \frac{\partial ^2u_{s,x}}{\partial x\, \partial \xi }\right)+\varepsilon  q_p q_{\lambda }^5 \left(-2 \theta  \frac{\partial ^3u_{s,x}}{\partial x^2\, \partial \tau }-\frac{\partial ^2u_{s,x}}{\partial x^2}\right)+q_p^3 q_{\lambda }^2 \left(u_{a,x} \frac{\partial ^2u_{a,x}}{\partial \xi \, \partial \tau }-\rho _a \frac{\partial ^2u_{a,x}}{\partial \xi ^2}-u_{a,x} \frac{\partial ^2\rho _a}{\partial \xi ^2}+\rho _a \frac{\partial ^2u_{a,x}}{\partial \tau ^2}-2 \frac{\partial u_{a,x}}{\partial \xi } \frac{\partial \rho _a}{\partial \xi }\right)+\\
                q_p^3 q_{\lambda }^3 \left(-2 \rho _a \frac{\partial ^2u_{a,x}}{\partial x\, \partial \xi }-2 u_{a,x} \frac{\partial ^2\rho _a}{\partial x\, \partial \xi }+u_{a,x} \frac{\partial ^2u_{a,x}}{\partial x\, \partial \tau }-2 \frac{\partial u_{a,x}}{\partial x} \frac{\partial \rho _a}{\partial \xi }-2 \frac{\partial u_{a,x}}{\partial \xi } \frac{\partial \rho _a}{\partial x}\right)+q_p^3 q_{\lambda }^4 \left(-\rho _a \frac{\partial ^2u_{a,x}}{\partial x^2}-u_{a,x} \frac{\partial ^2\rho _a}{\partial x^2}-2 \frac{\partial u_{a,x}}{\partial x} \frac{\partial \rho _a}{\partial x}\right)+\\
                \varepsilon  q_p^2 q_{\lambda }^2 \left(u_{a,x} \frac{\partial ^2u_{s,x}}{\partial \xi \, \partial \tau }+u_{s,x} \frac{\partial ^2u_{a,x}}{\partial \xi \, \partial \tau }-\rho _a \frac{\partial ^2u_{s,x}}{\partial \xi ^2}-\rho _s \frac{\partial ^2u_{a,x}}{\partial \xi ^2}-u_{a,x} \frac{\partial ^2\rho _s}{\partial \xi ^2}-u_{s,x} \frac{\partial ^2\rho _a}{\partial \xi ^2}+\rho _a \frac{\partial ^2u_{s,x}}{\partial \tau ^2}+\rho _s \frac{\partial ^2u_{a,x}}{\partial \tau ^2}-2 \frac{\partial u_{a,x}}{\partial \xi } \frac{\partial \rho _s}{\partial \xi }-2 \frac{\partial \rho _a}{\partial \xi } \frac{\partial u_{s,x}}{\partial \xi }\right)+\\
                \varepsilon  q_p^2 q_{\lambda }^3 \left(-2 \rho _a \frac{\partial ^2u_{s,x}}{\partial x\, \partial \xi }-2 \rho _s \frac{\partial ^2u_{a,x}}{\partial x\, \partial \xi }-2 u_{a,x} \frac{\partial ^2\rho _s}{\partial x\, \partial \xi }-2 u_{s,x} \frac{\partial ^2\rho _a}{\partial x\, \partial \xi }+u_{a,x} \frac{\partial ^2u_{s,x}}{\partial x\, \partial \tau }+u_{s,x} \frac{\partial ^2u_{a,x}}{\partial x\, \partial \tau }-2 \theta  \frac{\partial ^3u_{a,x}}{\partial \xi ^2\, \partial t}+2 \frac{\partial ^2u_{a,x}}{\partial t\, \partial \tau }-2 \frac{\partial u_{a,x}}{\partial x} \frac{\partial \rho _s}{\partial \xi }-2 \frac{\partial u_{a,x}}{\partial \xi } \frac{\partial \rho _s}{\partial x}-2 \frac{\partial \rho _a}{\partial \xi } \frac{\partial u_{s,x}}{\partial x}-2 \frac{\partial \rho _a}{\partial x} \frac{\partial u_{s,x}}{\partial \xi }-\frac{\partial F_x}{\partial t}\right)+\\
                \varepsilon  q_p^2 q_{\lambda }^4 \left(-4 \theta  \frac{\partial ^3u_{a,x}}{\partial x\, \partial \xi \, \partial t}-\rho _a \frac{\partial ^2u_{s,x}}{\partial x^2}-\rho _s \frac{\partial ^2u_{a,x}}{\partial x^2}-u_{a,x} \frac{\partial ^2\rho _s}{\partial x^2}-u_{s,x} \frac{\partial ^2\rho _a}{\partial x^2}-2 \frac{\partial u_{a,x}}{\partial x} \frac{\partial \rho _s}{\partial x}-2 \frac{\partial \rho _a}{\partial x} \frac{\partial u_{s,x}}{\partial x}\right)-2 \varepsilon  \theta  q_p^2 q_{\lambda }^5 \frac{\partial ^3u_{a,x}}{\partial x^2\, \partial t}+\\
                \varepsilon ^2 q_p q_{\lambda }^2 \left(u_{s,x} \frac{\partial ^2u_{s,x}}{\partial \xi \, \partial \tau }-\rho _s \frac{\partial ^2u_{s,x}}{\partial \xi ^2}-u_{s,x} \frac{\partial ^2\rho _s}{\partial \xi ^2}+\rho _s \frac{\partial ^2u_{s,x}}{\partial \tau ^2}-2 \frac{\partial u_{s,x}}{\partial \xi } \frac{\partial \rho _s}{\partial \xi }\right)+\varepsilon ^2 q_p q_{\lambda }^2 \left(u_{s,x} \frac{\partial ^2u_{s,x}}{\partial \xi \, \partial \tau }-\rho _s \frac{\partial ^2u_{s,x}}{\partial \xi ^2}-u_{s,x} \frac{\partial ^2\rho _s}{\partial \xi ^2}+\rho _s \frac{\partial ^2u_{s,x}}{\partial \tau ^2}-2 \frac{\partial u_{s,x}}{\partial \xi } \frac{\partial \rho _s}{\partial \xi }\right)+\\
                \varepsilon ^2 q_p q_{\lambda }^3 \left(-2 \theta  \frac{\partial ^3u_{s,x}}{\partial \xi ^2\, \partial t}-2 \rho _s \frac{\partial ^2u_{s,x}}{\partial x\, \partial \xi }-2 u_{s,x} \frac{\partial ^2\rho _s}{\partial x\, \partial \xi }+2 \frac{\partial ^2u_{s,x}}{\partial t\, \partial \tau }+u_{s,x} \frac{\partial ^2u_{s,x}}{\partial x\, \partial \tau }-2 \frac{\partial u_{s,x}}{\partial x} \frac{\partial \rho _s}{\partial \xi }-2 \frac{\partial u_{s,x}}{\partial \xi } \frac{\partial \rho _s}{\partial x}\right)+\varepsilon ^2 q_p q_{\lambda }^4 \left(-4 \theta  \frac{\partial ^3u_{s,x}}{\partial x\, \partial \xi \, \partial t}-\rho _s \frac{\partial ^2u_{s,x}}{\partial x^2}-u_{s,x} \frac{\partial ^2\rho _s}{\partial x^2}-2 \frac{\partial u_{s,x}}{\partial x} \frac{\partial \rho _s}{\partial x}\right)-\\
                2 \varepsilon ^2 \theta  q_p q_{\lambda }^5 \frac{\partial ^3u_{s,x}}{\partial x^2\, \partial t}+q_p^4 q_{\lambda } \left(u_{a,x} \rho _a \frac{\partial ^2u_{a,x}}{\partial \xi \, \partial \tau }-u_{a,x} \frac{\partial u_{a,x}}{\partial \tau } \frac{\partial \rho _a}{\partial \xi }-u_{a,x} \left(\frac{\partial u_{a,x}}{\partial \xi }\right)^2\right)+q_p^4 q_{\lambda }^2 \left(u_{a,x} \rho _a \frac{\partial ^2u_{a,x}}{\partial x\, \partial \tau }-u_{a,x} \frac{\partial u_{a,x}}{\partial \tau } \frac{\partial \rho _a}{\partial x}-2 u_{a,x} \frac{\partial u_{a,x}}{\partial x} \frac{\partial u_{a,x}}{\partial \xi }\right)+\\
                q_p^4 q_{\lambda }^3 u_{a,x} \left(-\left(\frac{\partial u_{a,x}}{\partial x}\right)^2\right)+\varepsilon  q_p^3 q_{\lambda } \left(u_{a,x} \rho _a \frac{\partial ^2u_{s,x}}{\partial \xi \, \partial \tau }+u_{a,x} \rho _s \frac{\partial ^2u_{a,x}}{\partial \xi \, \partial \tau }+\rho _a u_{s,x} \frac{\partial ^2u_{a,x}}{\partial \xi \, \partial \tau }-u_{a,x} \frac{\partial u_{a,x}}{\partial \tau } \frac{\partial \rho _s}{\partial \xi }-u_{a,x} \frac{\partial \rho _a}{\partial \xi } \frac{\partial u_{s,x}}{\partial \tau }-u_{s,x} \frac{\partial u_{a,x}}{\partial \tau } \frac{\partial \rho _a}{\partial \xi }-u_{s,x} \left(\frac{\partial u_{a,x}}{\partial \xi }\right)^2-2 u_{a,x} \frac{\partial u_{a,x}}{\partial \xi } \frac{\partial u_{s,x}}{\partial \xi }\right)+\\
                \varepsilon  q_p^3 q_{\lambda }^2 \left(u_{a,x} \rho _a \frac{\partial ^2u_{s,x}}{\partial x\, \partial \tau }+\rho _a u_{s,x} \frac{\partial ^2u_{a,x}}{\partial x\, \partial \tau }+u_{a,x} \rho _s \frac{\partial ^2u_{a,x}}{\partial x\, \partial \tau }+2 \rho _a \frac{\partial ^2u_{a,x}}{\partial t\, \partial \tau }+u_{a,x} \frac{\partial ^2u_{a,x}}{\partial \xi \, \partial t}-u_{a,x} \frac{\partial u_{a,x}}{\partial \tau } \frac{\partial \rho _s}{\partial x}-u_{a,x} \frac{\partial \rho _a}{\partial x} \frac{\partial u_{s,x}}{\partial \tau }-u_{s,x} \frac{\partial u_{a,x}}{\partial \tau } \frac{\partial \rho _a}{\partial x}-2 u_{a,x} \frac{\partial u_{a,x}}{\partial \xi } \frac{\partial u_{s,x}}{\partial x}-\right.\\
                \left.2 u_{a,x} \frac{\partial u_{a,x}}{\partial x} \frac{\partial u_{s,x}}{\partial \xi }-2 u_{s,x} \frac{\partial u_{a,x}}{\partial x} \frac{\partial u_{a,x}}{\partial \xi }\right)+\varepsilon  q_p^3 q_{\lambda }^3 \left(u_{a,x} \frac{\partial ^2u_{a,x}}{\partial x\, \partial t}-u_{s,x} \left(\frac{\partial u_{a,x}}{\partial x}\right)^2-2 u_{a,x} \frac{\partial u_{a,x}}{\partial x} \frac{\partial u_{s,x}}{\partial x}\right)+\\
                \varepsilon ^2 q_p^2 q_{\lambda } \left(u_{a,x} \rho _s \frac{\partial ^2u_{s,x}}{\partial \xi \, \partial \tau }+\rho _a u_{s,x} \frac{\partial ^2u_{s,x}}{\partial \xi \, \partial \tau }+u_{s,x} \rho _s \frac{\partial ^2u_{a,x}}{\partial \xi \, \partial \tau }-u_{a,x} \frac{\partial u_{s,x}}{\partial \tau } \frac{\partial \rho _s}{\partial \xi }-u_{s,x} \frac{\partial u_{a,x}}{\partial \tau } \frac{\partial \rho _s}{\partial \xi }-u_{s,x} \frac{\partial \rho _a}{\partial \xi } \frac{\partial u_{s,x}}{\partial \tau }-u_{a,x} \left(\frac{\partial u_{s,x}}{\partial \xi }\right)^2-2 u_{s,x} \frac{\partial u_{a,x}}{\partial \xi } \frac{\partial u_{s,x}}{\partial \xi }\right)+\\
                \varepsilon ^2 q_p^2 q_{\lambda }^2 \left(2 \rho _a \frac{\partial ^2u_{s,x}}{\partial t\, \partial \tau }+\rho _a u_{s,x} \frac{\partial ^2u_{s,x}}{\partial x\, \partial \tau }+2 \rho _s \frac{\partial ^2u_{a,x}}{\partial t\, \partial \tau }+u_{a,x} \rho _s \frac{\partial ^2u_{s,x}}{\partial x\, \partial \tau }+u_{s,x} \rho _s \frac{\partial ^2u_{a,x}}{\partial x\, \partial \tau }+u_{a,x} \frac{\partial ^2u_{s,x}}{\partial \xi \, \partial t}+u_{s,x} \frac{\partial ^2u_{a,x}}{\partial \xi \, \partial t}-u_{a,x} \frac{\partial u_{s,x}}{\partial \tau } \frac{\partial \rho _s}{\partial x}-u_{s,x} \frac{\partial u_{a,x}}{\partial \tau } \frac{\partial \rho _s}{\partial x}-u_{s,x} \frac{\partial \rho _a}{\partial x} \frac{\partial u_{s,x}}{\partial \tau }-\right.\\
                \left. 2 u_{a,x} \frac{\partial u_{s,x}}{\partial x} \frac{\partial u_{s,x}}{\partial \xi }-2 u_{s,x} \frac{\partial u_{a,x}}{\partial \xi } \frac{\partial u_{s,x}}{\partial x}-2 u_{s,x} \frac{\partial u_{a,x}}{\partial x} \frac{\partial u_{s,x}}{\partial \xi }\right)+\\
                \varepsilon ^2 q_p^2 q_{\lambda }^3 \left(u_{a,x} \frac{\partial ^2u_{s,x}}{\partial x\, \partial t}+u_{s,x} \frac{\partial ^2u_{a,x}}{\partial x\, \partial t}-u_{a,x} \left(\frac{\partial u_{s,x}}{\partial x}\right)^2-2 u_{s,x} \frac{\partial u_{a,x}}{\partial x} \frac{\partial u_{s,x}}{\partial x}+\frac{\partial ^2u_{a,x}}{\partial t^2}\right)+\varepsilon ^3 q_p q_{\lambda } \left(u_{s,x} \rho _s \frac{\partial ^2u_{s,x}}{\partial \xi \, \partial \tau }-u_{s,x} \frac{\partial u_{s,x}}{\partial \tau } \frac{\partial \rho _s}{\partial \xi }-u_{s,x} \left(\frac{\partial u_{s,x}}{\partial \xi }\right)^2\right)+\\
                \varepsilon ^3 q_p q_{\lambda }^2 \left(2 \rho _s \frac{\partial ^2u_{s,x}}{\partial t\, \partial \tau }+u_{s,x} \rho _s \frac{\partial ^2u_{s,x}}{\partial x\, \partial \tau }+u_{s,x} \frac{\partial ^2u_{s,x}}{\partial \xi \, \partial t}-u_{s,x} \frac{\partial u_{s,x}}{\partial \tau } \frac{\partial \rho _s}{\partial x}-2 u_{s,x} \frac{\partial u_{s,x}}{\partial x} \frac{\partial u_{s,x}}{\partial \xi }\right)+\\
                \varepsilon ^3 q_p q_{\lambda }^3 \left(u_{s,x} \frac{\partial ^2u_{s,x}}{\partial x\, \partial t}+\frac{\partial ^2u_{s,x}}{\partial t^2}-u_{s,x} \left(\frac{\partial u_{s,x}}{\partial x}\right)^2\right)+q_p^5 \left(u_{a,x} \left(-\rho _a\right) \left(\frac{\partial u_{a,x}}{\partial \xi }\right)^2-u_{a,x}^2 \frac{\partial u_{a,x}}{\partial \xi } \frac{\partial \rho _a}{\partial \xi }\right)+q_p^5 q_{\lambda } \left(u_{a,x}^2 \frac{\partial u_{a,x}}{\partial x} \left(-\frac{\partial \rho _a}{\partial \xi }\right)-u_{a,x}^2 \frac{\partial u_{a,x}}{\partial \xi } \frac{\partial \rho _a}{\partial x}-2 u_{a,x} \rho _a \frac{\partial u_{a,x}}{\partial x} \frac{\partial u_{a,x}}{\partial \xi }\right)+\\
                q_p^5 q_{\lambda }^2 \left(u_{a,x} \left(-\rho _a\right) \left(\frac{\partial u_{a,x}}{\partial x}\right)^2-u_{a,x}^2 \frac{\partial u_{a,x}}{\partial x} \frac{\partial \rho _a}{\partial x}\right)+\varepsilon  q_p^4 \left(u_{a,x} \left(-\rho _s\right) \left(\frac{\partial u_{a,x}}{\partial \xi }\right)^2-\rho _a u_{s,x} \left(\frac{\partial u_{a,x}}{\partial \xi }\right)^2-u_{a,x}^2 \frac{\partial u_{a,x}}{\partial \xi } \frac{\partial \rho _s}{\partial \xi }-2 u_{a,x} \rho _a \frac{\partial u_{a,x}}{\partial \xi } \frac{\partial u_{s,x}}{\partial \xi }-2 u_{a,x} u_{s,x} \frac{\partial u_{a,x}}{\partial \xi } \frac{\partial \rho _a}{\partial \xi }-u_{a,x}^2 \frac{\partial \rho _a}{\partial \xi } \frac{\partial u_{s,x}}{\partial \xi }\right)+\\
                \varepsilon  q_p^4 q_{\lambda } \left(u_{a,x} \rho _a \frac{\partial ^2u_{a,x}}{\partial \xi \, \partial t}+u_{a,x}^2 \frac{\partial u_{a,x}}{\partial x} \left(-\frac{\partial \rho _s}{\partial \xi }\right)-u_{a,x}^2 \frac{\partial u_{a,x}}{\partial \xi } \frac{\partial \rho _s}{\partial x}-u_{a,x}^2 \frac{\partial \rho _a}{\partial \xi } \frac{\partial u_{s,x}}{\partial x}-u_{a,x}^2 \frac{\partial \rho _a}{\partial x} \frac{\partial u_{s,x}}{\partial \xi }-2 u_{a,x} \rho _a \frac{\partial u_{a,x}}{\partial \xi } \frac{\partial u_{s,x}}{\partial x}-2 u_{a,x} \rho _a \frac{\partial u_{a,x}}{\partial x} \frac{\partial u_{s,x}}{\partial \xi }-2 u_{a,x} \rho _s \frac{\partial u_{a,x}}{\partial x} \frac{\partial u_{a,x}}{\partial \xi }-\right.\\
                \left. 2 u_{a,x} u_{s,x} \frac{\partial u_{a,x}}{\partial x} \frac{\partial \rho _a}{\partial \xi }-2 u_{a,x} u_{s,x} \frac{\partial u_{a,x}}{\partial \xi } \frac{\partial \rho _a}{\partial x}-2 \rho _a u_{s,x} \frac{\partial u_{a,x}}{\partial x} \frac{\partial u_{a,x}}{\partial \xi }-u_{a,x} \frac{\partial u_{a,x}}{\partial t} \frac{\partial \rho _a}{\partial \xi }\right)+\\
                \varepsilon  q_p^4 q_{\lambda }^2 \left(u_{a,x} \rho _a \frac{\partial ^2u_{a,x}}{\partial x\, \partial t}+u_{a,x} \left(-\rho _s\right) \left(\frac{\partial u_{a,x}}{\partial x}\right)^2-\rho _a u_{s,x} \left(\frac{\partial u_{a,x}}{\partial x}\right)^2-u_{a,x}^2 \frac{\partial u_{a,x}}{\partial x} \frac{\partial \rho _s}{\partial x}-2 u_{a,x} \rho _a \frac{\partial u_{a,x}}{\partial x} \frac{\partial u_{s,x}}{\partial x}-2 u_{a,x} u_{s,x} \frac{\partial u_{a,x}}{\partial x} \frac{\partial \rho _a}{\partial x}-u_{a,x}^2 \frac{\partial \rho _a}{\partial x} \frac{\partial u_{s,x}}{\partial x}-u_{a,x} \frac{\partial u_{a,x}}{\partial t} \frac{\partial \rho _a}{\partial x}\right)+\\
                \varepsilon ^2 q_p^3 \left(u_{s,x} \left(-\rho _s\right) \left(\frac{\partial u_{a,x}}{\partial \xi }\right)^2-u_{s,x}^2 \frac{\partial u_{a,x}}{\partial \xi } \frac{\partial \rho _a}{\partial \xi }-2 u_{a,x} \rho _s \frac{\partial u_{a,x}}{\partial \xi } \frac{\partial u_{s,x}}{\partial \xi }-2 \rho _a u_{s,x} \frac{\partial u_{a,x}}{\partial \xi } \frac{\partial u_{s,x}}{\partial \xi }-2 u_{a,x} u_{s,x} \frac{\partial u_{a,x}}{\partial \xi } \frac{\partial \rho _s}{\partial \xi }-u_{a,x}^2 \frac{\partial u_{s,x}}{\partial \xi } \frac{\partial \rho _s}{\partial \xi }-u_{a,x} \rho _a \left(\frac{\partial u_{s,x}}{\partial \xi }\right)^2-2 u_{a,x} u_{s,x} \frac{\partial \rho _a}{\partial \xi } \frac{\partial u_{s,x}}{\partial \xi }\right)+\\
                \varepsilon ^2 q_p^3 q_{\lambda } \left(u_{a,x} \rho _a \frac{\partial ^2u_{s,x}}{\partial \xi \, \partial t}+u_{a,x} \rho _s \frac{\partial ^2u_{a,x}}{\partial \xi \, \partial t}+\rho _a u_{s,x} \frac{\partial ^2u_{a,x}}{\partial \xi \, \partial t}+u_{a,x}^2 \frac{\partial u_{s,x}}{\partial x} \left(-\frac{\partial \rho _s}{\partial \xi }\right)-u_{a,x}^2 \frac{\partial u_{s,x}}{\partial \xi } \frac{\partial \rho _s}{\partial x}-u_{a,x} \frac{\partial u_{a,x}}{\partial t} \frac{\partial \rho _s}{\partial \xi }-u_{a,x} \frac{\partial \rho _a}{\partial \xi } \frac{\partial u_{s,x}}{\partial t}-\right.\\
                \left. 2 u_{a,x} \rho _a \frac{\partial u_{s,x}}{\partial x} \frac{\partial u_{s,x}}{\partial \xi }-2 u_{a,x} \rho _s \frac{\partial u_{a,x}}{\partial \xi } \frac{\partial u_{s,x}}{\partial x}-2 u_{a,x} \rho _s \frac{\partial u_{a,x}}{\partial x} \frac{\partial u_{s,x}}{\partial \xi }-2 u_{a,x} u_{s,x} \frac{\partial u_{a,x}}{\partial x} \frac{\partial \rho _s}{\partial \xi }-2 u_{a,x} u_{s,x} \frac{\partial u_{a,x}}{\partial \xi } \frac{\partial \rho _s}{\partial x}-2 u_{a,x} u_{s,x} \frac{\partial \rho _a}{\partial \xi } \frac{\partial u_{s,x}}{\partial x}-\right.\\
                \left. 2 u_{a,x} u_{s,x} \frac{\partial \rho _a}{\partial x} \frac{\partial u_{s,x}}{\partial \xi }-u_{s,x}^2 \frac{\partial u_{a,x}}{\partial x} \frac{\partial \rho _a}{\partial \xi }-u_{s,x}^2 \frac{\partial u_{a,x}}{\partial \xi } \frac{\partial \rho _a}{\partial x}-u_{s,x} \frac{\partial u_{a,x}}{\partial t} \frac{\partial \rho _a}{\partial \xi }-2 \rho _a u_{s,x} \frac{\partial u_{a,x}}{\partial \xi } \frac{\partial u_{s,x}}{\partial x}-2 \rho _a u_{s,x} \frac{\partial u_{a,x}}{\partial x} \frac{\partial u_{s,x}}{\partial \xi }-2 u_{s,x} \rho _s \frac{\partial u_{a,x}}{\partial x} \frac{\partial u_{a,x}}{\partial \xi }\right)+\\
                \varepsilon ^2 q_p^3 q_{\lambda }^2 \left(u_{a,x} \rho _a \frac{\partial ^2u_{s,x}}{\partial x\, \partial t}+u_{a,x} \rho _s \frac{\partial ^2u_{a,x}}{\partial x\, \partial t}+\rho _a u_{s,x} \frac{\partial ^2u_{a,x}}{\partial x\, \partial t}-u_{s,x}\rho _s \left(\frac{\partial u_{a,x}}{\partial x}\right)^2-u_{s,x}^2 \frac{\partial u_{a,x}}{\partial x} \frac{\partial \rho _a}{\partial x}-2 u_{a,x} \rho _s \frac{\partial u_{a,x}}{\partial x} \frac{\partial u_{s,x}}{\partial x}-\right.\\
                \left. 2 \rho _a u_{s,x} \frac{\partial u_{a,x}}{\partial x} \frac{\partial u_{s,x}}{\partial x}-2 u_{a,x} u_{s,x} \frac{\partial u_{a,x}}{\partial x} \frac{\partial \rho _s}{\partial x}-u_{a,x}^2 \frac{\partial u_{s,x}}{\partial x} \frac{\partial \rho _s}{\partial x}-u_{a,x} \frac{\partial u_{a,x}}{\partial t} \frac{\partial \rho _s}{\partial x}-u_{a,x} \frac{\partial \rho _a}{\partial x} \frac{\partial u_{s,x}}{\partial t}-u_{a,x} \rho _a \left(\frac{\partial u_{s,x}}{\partial x}\right)^2-u_{s,x} \frac{\partial u_{a,x}}{\partial t} \frac{\partial \rho _a}{\partial x}-2 u_{a,x} u_{s,x} \frac{\partial \rho _a}{\partial x} \frac{\partial u_{s,x}}{\partial x}+\rho _a \frac{\partial ^2u_{a,x}}{\partial t^2}\right)+\\
                \varepsilon ^3 q_p^2 \left(u_{a,x} \left(-\rho _s\right) \left(\frac{\partial u_{s,x}}{\partial \xi }\right)^2-\rho _a u_{s,x} \left(\frac{\partial u_{s,x}}{\partial \xi }\right)^2-u_{s,x}^2 \frac{\partial \rho _a}{\partial \xi } \frac{\partial u_{s,x}}{\partial \xi }-2 u_{s,x} \rho _s \frac{\partial u_{a,x}}{\partial \xi } \frac{\partial u_{s,x}}{\partial \xi }-2 u_{a,x} u_{s,x} \frac{\partial u_{s,x}}{\partial \xi } \frac{\partial \rho _s}{\partial \xi }-u_{s,x}^2 \frac{\partial u_{a,x}}{\partial \xi } \frac{\partial \rho _s}{\partial \xi }\right)+\\
                \varepsilon ^3 q_p^2 q_{\lambda } \left(\rho _a u_{s,x} \frac{\partial ^2u_{s,x}}{\partial \xi \, \partial t}+u_{s,x} \rho _s \frac{\partial ^2u_{a,x}}{\partial \xi \, \partial t}+u_{a,x} \rho _s \frac{\partial ^2u_{s,x}}{\partial \xi \, \partial t}+u_{s,x}^2 \frac{\partial u_{a,x}}{\partial x} \left(-\frac{\partial \rho _s}{\partial \xi }\right)-u_{s,x}^2 \frac{\partial u_{a,x}}{\partial \xi } \frac{\partial \rho _s}{\partial x}-u_{s,x}^2 \frac{\partial \rho _a}{\partial \xi } \frac{\partial u_{s,x}}{\partial x}-u_{s,x}^2 \frac{\partial \rho _a}{\partial x} \frac{\partial u_{s,x}}{\partial \xi }-u_{s,x} \frac{\partial u_{a,x}}{\partial t} \frac{\partial \rho _s}{\partial \xi }-\right.\\
                \left. u_{s,x} \frac{\partial \rho _a}{\partial \xi } \frac{\partial u_{s,x}}{\partial t}-2 \rho _a u_{s,x} \frac{\partial u_{s,x}}{\partial x} \frac{\partial u_{s,x}}{\partial \xi }-2 u_{s,x} \rho _s \frac{\partial u_{a,x}}{\partial \xi } \frac{\partial u_{s,x}}{\partial x}-2 u_{s,x} \rho _s \frac{\partial u_{a,x}}{\partial x} \frac{\partial u_{s,x}}{\partial \xi }-2 u_{a,x} u_{s,x} \frac{\partial u_{s,x}}{\partial x} \frac{\partial \rho _s}{\partial \xi }-2 u_{a,x} u_{s,x} \frac{\partial u_{s,x}}{\partial \xi } \frac{\partial \rho _s}{\partial x}-u_{a,x} \frac{\partial u_{s,x}}{\partial t} \frac{\partial \rho _s}{\partial \xi }-2 u_{a,x} \rho _s \frac{\partial u_{s,x}}{\partial x} \frac{\partial u_{s,x}}{\partial \xi }\right)+\\
                \varepsilon ^3 q_p^2 q_{\lambda }^2 \left(u_{a,x} \rho _s \frac{\partial ^2u_{s,x}}{\partial x\, \partial t}+\rho _a u_{s,x} \frac{\partial ^2u_{s,x}}{\partial x\, \partial t}+u_{s,x} \rho _s \frac{\partial ^2u_{a,x}}{\partial x\, \partial t}+\rho _a \frac{\partial ^2u_{s,x}}{\partial t^2}+\rho _s \frac{\partial ^2u_{a,x}}{\partial t^2}+u_{a,x} \left(-\rho _s\right) \left(\frac{\partial u_{s,x}}{\partial x}\right)^2-\rho _a u_{s,x} \left(\frac{\partial u_{s,x}}{\partial x}\right)^2-u_{s,x}^2 \frac{\partial \rho _a}{\partial x} \frac{\partial u_{s,x}}{\partial x}-\right.\\
                \left. 2 u_{s,x} \rho _s \frac{\partial u_{a,x}}{\partial x} \frac{\partial u_{s,x}}{\partial x}-2 u_{a,x} u_{s,x} \frac{\partial u_{s,x}}{\partial x} \frac{\partial \rho _s}{\partial x}-u_{s,x}^2 \frac{\partial u_{a,x}}{\partial x} \frac{\partial \rho _s}{\partial x}-u_{a,x} \frac{\partial u_{s,x}}{\partial t} \frac{\partial \rho _s}{\partial x}-u_{s,x} \frac{\partial u_{a,x}}{\partial t} \frac{\partial \rho _s}{\partial x}-u_{s,x} \frac{\partial \rho _a}{\partial x} \frac{\partial u_{s,x}}{\partial t}\right)+\\
                \varepsilon ^4 q_p \left(u_{s,x} \left(-\rho _s\right) \left(\frac{\partial u_{s,x}}{\partial \xi }\right)^2-u_{s,x}^2 \frac{\partial u_{s,x}}{\partial \xi } \frac{\partial \rho _s}{\partial \xi }\right)+\varepsilon ^4 q_p q_{\lambda } \left(u_{s,x} \rho _s \frac{\partial ^2u_{s,x}}{\partial \xi \, \partial t}+u_{s,x}^2 \frac{\partial u_{s,x}}{\partial x} \left(-\frac{\partial \rho _s}{\partial \xi }\right)-u_{s,x}^2 \frac{\partial u_{s,x}}{\partial \xi } \frac{\partial \rho _s}{\partial x}-u_{s,x} \frac{\partial u_{s,x}}{\partial t} \frac{\partial \rho _s}{\partial \xi }-2 u_{s,x} \rho _s \frac{\partial u_{s,x}}{\partial x} \frac{\partial u_{s,x}}{\partial \xi }\right)+\\
                \varepsilon ^4 q_p q_{\lambda }^2 \left(u_{s,x} \rho _s \frac{\partial ^2u_{s,x}}{\partial x\, \partial t}+\rho _s \frac{\partial ^2u_{s,x}}{\partial t^2}+u_{s,x} \left(-\rho _s\right) \left(\frac{\partial u_{s,x}}{\partial x}\right)^2-u_{s,x}^2 \frac{\partial u_{s,x}}{\partial x} \frac{\partial \rho _s}{\partial x}-u_{s,x} \frac{\partial u_{s,x}}{\partial t} \frac{\partial \rho _s}{\partial x}\right)=0.
                \end{multline}
              \end{minipage}
            }
            \end{widetext}

After normalizing by the leading order magnitude, $q_p^2\,q_{\lambda}^3$, eqn.\,\eqref{eq:266term} is written
%
\begin{align}
	\mathcal{A}\,u^{(a)}_x = \partial_{\tau}F_x+\mathcal{O}\left[q_{\lambda}\right],
\end{align}
where the linear field operator $\mathcal{A}=\partial_{\tau}^2-\partial_{\xi}^2-2\,\theta\,\partial_{\tau}\partial_{\xi}^2$ describes the acoustics. By discarding terms less than $\mathcal{O}(1)$ and setting the force $F_x$ equal to zero, we arrive at
\begin{equation}
\mathcal{A}\,u^{(a)}_x=0.\label{eqn:homo_acoustic_eqn}
\end{equation} 
This leading order equation is a pervasive result~\citep{rudenko_theoretical_1977,shutilov_fundamental_1988,riley_steady_2001-1,riaud_influence_2017} describing damped propagation of a linear acoustic wave in a dissipative medium. We follow \citet{riley_steady_2001-1} in retaining the $2\theta\partial_\tau\partial_\xi^2$ term, because even if the term is made small by a small value of $\theta$, the term is responsible for attenuation, which is required to simultaneously satisfy both boundary conditions.

The stationary solution to the damped wave equation $\mathcal{A}\,u^{(a)}_x=0$, representing an acoustic wave generated from the vibrating origin of a semiinfinite domain, is
%
\begin{align}\label{eq:stationary_acoustic_wave}
	\lim_{\tau\rightarrow\tau_{\infty}^{-}}u^{(a)}_x\approx\exp[\iota(\kappa\,\xi-\tau)],
\end{align}
where $\kappa = \kappa_r+\iota\,\kappa_i$ with
%
\begin{subequations}
\begin{align}
	\kappa_r = \sqrt{\frac{\sqrt{1+4\,\theta^2}+1}{2(1+4\,\theta^2)}}, \\
	\kappa_i = \sqrt{\frac{\sqrt{1+4\,\theta^2}-1}{2(1+4\,\theta^2)}},
\end{align}\label{eqn:kappavars}
\end{subequations}
and where $\kappa_i\approx\theta \ll 1$ and $\kappa_r\approx1$ in water for frequencies relevant to this analysis (\emph{see} these quantities plotted in Fig.~\ref{fig:f01_system_scales}). This is a substantially different result than that obtained by \citet{riley_steady_2001-1}. In Riley's study, there may be a mistake, as the solution given in eqn.\,(9) does not satisfy the first-order damped equation obtained from eqn.\,(8).

\subsubsection{Transient Burgers streaming}
We now return to eqn.\,\eqref{eq:nondim_ns_equations} to deduce the equation for the streaming that arises from the acoustic wave generated in the semiinfinite domain presented in eqns.\,\eqref{eq:stationary_acoustic_wave}~and~\eqref{eqn:kappavars}. Substituting the partitioning equations~\eqref{eq:super_nondim}~and~\eqref{eq:modified_density_partition} for $u_x$ and $\rho$ and expanding as in the previous section leads to another finite ($47$-term) expression given by

\begin{widetext}
\resizebox{0.7\linewidth}{!}{
\begin{minipage}{\linewidth}
    \begin{multline}\label{eqn:47term}
        q_p q_{\lambda }^2 \left(\frac{\partial u_{a,x}}{\partial \tau }+\frac{\partial \rho _a}{\partial \xi }-F_x\right)+q_p q_{\lambda }^3 \left(\frac{\partial \rho _a}{\partial x}-2 \theta  \frac{\partial ^2u_{a,x}}{\partial \xi ^2}\right)-4 \theta  q_p q_{\lambda }^4 \frac{\partial ^2u_{a,x}}{\partial x\, \partial \xi }-2 \theta  q_p q_{\lambda }^5 \frac{\partial ^2u_{a,x}}{\partial x^2}+\varepsilon  q_{\lambda }^2 \left(\frac{\partial u_{s,x}}{\partial \tau }+\frac{\partial \rho _s}{\partial \xi }\right)+\varepsilon  q_{\lambda }^3 \left(\frac{\partial \rho _s}{\partial x}-2 \theta  \frac{\partial ^2u_{s,x}}{\partial \xi ^2}\right)-4 \varepsilon  \theta  q_{\lambda }^4 \frac{\partial ^2u_{s,x}}{\partial x\, \partial \xi }-2 \varepsilon  \theta  q_{\lambda }^5 \frac{\partial ^2u_{s,x}}{\partial x^2}+\\
        q_p^2 q_{\lambda } \left(\rho _a \frac{\partial u_{a,x}}{\partial \tau }+u_{a,x} \frac{\partial u_{a,x}}{\partial \xi }\right)+q_p^2 q_{\lambda }^2 u_{a,x} \frac{\partial u_{a,x}}{\partial x}+\varepsilon  q_p q_{\lambda } \left(\rho _a \frac{\partial u_{s,x}}{\partial \tau }+\rho _s \frac{\partial u_{a,x}}{\partial \tau }+u_{a,x} \frac{\partial u_{s,x}}{\partial \xi }+u_{s,x} \frac{\partial u_{a,x}}{\partial \xi }\right)+\varepsilon  q_p q_{\lambda }^2 \left(u_{a,x} \frac{\partial u_{s,x}}{\partial x}+u_{s,x} \frac{\partial u_{a,x}}{\partial x}+\frac{\partial u_{a,x}}{\partial t}\right)+\\
        \varepsilon ^2 q_{\lambda } \left(\rho _s \frac{\partial u_{s,x}}{\partial \tau }+u_{s,x} \frac{\partial u_{s,x}}{\partial \xi }\right)+\varepsilon ^2 q_{\lambda }^2 \left(\frac{\partial u_{s,x}}{\partial t}+u_{s,x} \frac{\partial u_{s,x}}{\partial x}\right)+q_p^3 u_{a,x} \rho _a \frac{\partial u_{a,x}}{\partial \xi }+q_p^3 q_{\lambda } u_{a,x} \rho _a \frac{\partial u_{a,x}}{\partial x}+\varepsilon  q_p^2 \left(u_{a,x} \rho _a \frac{\partial u_{s,x}}{\partial \xi }+u_{a,x} \rho _s \frac{\partial u_{a,x}}{\partial \xi }+\rho _a u_{s,x} \frac{\partial u_{a,x}}{\partial \xi }\right)+\\
        \varepsilon  q_p^2 q_{\lambda } \left(u_{a,x} \rho _a \frac{\partial u_{s,x}}{\partial x}+\rho _a u_{s,x} \frac{\partial u_{a,x}}{\partial x}+u_{a,x} \rho _s \frac{\partial u_{a,x}}{\partial x}+\rho _a \frac{\partial u_{a,x}}{\partial t}\right)+\varepsilon ^2 q_p \left(u_{a,x} \rho _s \frac{\partial u_{s,x}}{\partial \xi }+\rho _a u_{s,x} \frac{\partial u_{s,x}}{\partial \xi }+u_{s,x} \rho _s \frac{\partial u_{a,x}}{\partial \xi }\right)+\\
        \varepsilon ^2 q_p q_{\lambda } \left(\rho _a \frac{\partial u_{s,x}}{\partial t}+\rho _a u_{s,x} \frac{\partial u_{s,x}}{\partial x}+\rho _s \frac{\partial u_{a,x}}{\partial t}+u_{a,x} \rho _s \frac{\partial u_{s,x}}{\partial x}+u_{s,x} \rho _s \frac{\partial u_{a,x}}{\partial x}\right)+\varepsilon ^3 u_{s,x} \rho _s \frac{\partial u_{s,x}}{\partial \xi }+\varepsilon ^3 q_{\lambda } \left(\rho _s \frac{\partial u_{s,x}}{\partial t}+u_{s,x} \rho _s \frac{\partial u_{s,x}}{\partial x}\right)=0.
	\end{multline}
\end{minipage}
}
\end{widetext}

By again setting the applied force $F_x = 0$ and then applying the differential and integral constraints described back in subsection~\ref{subsec:constraints}, the expression is reduced to seventeen terms. The remaining seventeen terms are

\begin{widetext}
\resizebox{0.7\linewidth}{!}{
\begin{minipage}{\linewidth}
    \begin{multline}\label{eqn:17term}
        \varepsilon  q_{\lambda }^3 \frac{\partial \rho _s}{\partial x}-2 \varepsilon  \theta  q_{\lambda }^5 \frac{\partial ^2u_{s,x}}{\partial x^2}+q_p^2 q_{\lambda } \left(\rho _a \frac{\partial u_{a,x}}{\partial \tau }+u_{a,x} \frac{\partial u_{a,x}}{\partial \xi }\right)+\varepsilon  q_p q_{\lambda } \left(\rho _s \frac{\partial u_{a,x}}{\partial \tau }+u_{s,x} \frac{\partial u_{a,x}}{\partial \xi }\right)+q_p^2 q_{\lambda }^2 u_{a,x} \frac{\partial u_{a,x}}{\partial x}+\varepsilon ^2 q_{\lambda }^2 \left(\frac{\partial u_{s,x}}{\partial t}+u_{s,x} \frac{\partial u_{s,x}}{\partial x}\right)+q_p^3 u_{a,x} \rho _a \frac{\partial u_{a,x}}{\partial \xi }+\\
        q_p^3 q_{\lambda } u_{a,x} \rho _a \frac{\partial u_{a,x}}{\partial x}+\varepsilon  q_p^2 \left(\rho _s u_{a,x} \frac{\partial u_{a,x}}{\partial \xi }+u_{s,x} \rho _a \frac{\partial u_{a,x}}{\partial \xi }\right)+\varepsilon  q_p^2 q_{\lambda } \left(\frac{\partial u_{s,x}}{\partial x} u_{a,x} \rho _a+u_{s,x} \rho _a \frac{\partial u_{a,x}}{\partial x}+\rho _s u_{a,x} \frac{\partial u_{a,x}}{\partial x}+\rho _a \frac{\partial u_{a,x}}{\partial t}\right)+\varepsilon ^3 q_{\lambda } \left(\rho _s \frac{\partial u_{s,x}}{\partial t}+u_{s,x} \rho _s \frac{\partial u_{s,x}}{\partial x}\right)=0.
	\end{multline}
\end{minipage}
}
\end{widetext}

Of these remaining seventeen terms, we discard those terms of third or higher order in combinations of $q_p$ and $S^{-1}$, leaving seven terms to consider:

\begin{widetext}
\begin{minipage}{\linewidth}
    \begin{multline}\label{eqn:5termeqn}
        \varepsilon  q_{\lambda }^3 \frac{\partial \rho _s}{\partial x}-2 \varepsilon  \theta  q_{\lambda }^5 \frac{\partial ^2u_{s,x}}{\partial x^2}+q_p^2 q_{\lambda } \left(\rho _a \frac{\partial u_{a,x}}{\partial \tau }+u_{a,x} \frac{\partial u_{a,x}}{\partial \xi }\right)+q_p^2 q_{\lambda }^2 u_{a,x} \frac{\partial u_{a,x}}{\partial x}+\varepsilon ^2 q_{\lambda }^2 \left(\frac{\partial u_{s,x}}{\partial t}+u_{s,x} \frac{\partial u_{s,x}}{\partial x}\right)=0.
	\end{multline}
\end{minipage}
\end{widetext}

The leading order $\varepsilon$ terms are
\begin{equation}
S^{-1}\,q_{\lambda}^3\,\partial_x\,\rho^{(s)} \quad\text{at}\quad \mathcal{O}\left[q_{\lambda}^3\right]
\label{eqn:streamingdensityconstraint}
\end{equation}
and 
\begin{equation}
2\,S^{-1}q_{\lambda}^5\,\theta\,\partial_x^2\,u^{(s)}_x \quad\text{at}\quad \mathcal{O}\left[q_{\lambda}^5\right].\label{eqn:constraintforlater}
\end{equation}

Solving eqn.\,\eqref{eqn:streamingdensityconstraint} at leading order implies solving
	%
	\begin{align}
		\partial_x\,\rho^{(s)} \approx 0,
	\end{align}
so that, after applying a homogeneous source condition (where a ``homogeneous'' boundary condition means that it is zero), we have $\rho^{(s)}(x,t)\approx0$ for all 
$\{x,t\}$. In other words, elimination of this term from eqn.\,\eqref{eqn:47term} implies that the changes in the density to the streaming flow are negligible, that the streaming flow itself is incompressible. Traditionally, this is assumed prior to derivation without rigorous justification. By contrast, here the result arises naturally from a term in eqn.\,\eqref{eqn:47term}, eqn.\,\eqref{eqn:streamingdensityconstraint} being solved at $\mathcal{O}[\varepsilon\,q_{\lambda}^3]$: $S^{-1}\,q_{\lambda}^3\,\partial_x\,\rho^{(s)}=0$.

Moving now to those terms present at first order in  $q_{\lambda}$ and second order in combinations of $S^{-1}$ and $q_p$, we have
%
\begin{align}\label{eq:zero_lagrangian_identity}
	\langle u^{(a)}_x\partial_{\xi}u^{(a)}_x\rangle_{\xi,\tau} +\langle\rho^{(a)}\,\partial_{\tau}u^{(a)}_x\rangle_{\xi,\tau}=0.
\end{align}
This is an interesting departure from past analyses~\citep{lighthill_acoustic_1978,riaud_influence_2017}, where one defines 
%
\begin{align}
	\langle\partial_{\xi}\mathcal{L}\rangle_{\xi,\tau}=\langle \partial_{\xi}\,\mathcal{T}\rangle_{\xi,\tau} -\langle\partial_{\xi}\,\mathcal{U}\rangle_{\xi,\tau},
	\label{eqn:classicacousticlagrangian}
\end{align}
in terms of the acoustic Lagrangian $\mathcal{L}=\mathcal{T}-\mathcal{U}$, where $\partial_{\xi}\,\mathcal{T}=u^{(a)}_x\partial_{\xi}u^{(a)}_x$ and $\partial_{\xi}\,\mathcal{U}=-\rho^{(a)}\,\partial_{\tau}u^{(a)}_x$ are the gradient of the kinetic and potential acoustic energies, respectively. The acoustic density can be written in terms of the acoustic velocity by solving the leading order equation in an expansion of the continuity equation [see supplemental information]:
\begin{align}
    \rho^{(a)}=-\int^t_0\partial_{\xi}u^{(a)}\,d\tau,
\end{align}
so that $\partial_{\xi}\mathcal{U}=\partial_{\tau}u^{(a)}_x\,\int^t_0\partial_{\xi}u^{(a)}\,d\tau$. Then with eqn.\,\eqref{eq:stationary_acoustic_wave} in eqn.\,\eqref{eqn:classicacousticlagrangian},
%
\begin{align}
	\begin{split}
		\max_{\xi\geq\pi/\kappa_r}|\langle\partial_{\xi}\mathcal{L}\rangle_{\xi,\tau}| &= \frac{\kappa_r}{2\,\pi}\E^{\frac{-2\,\pi\,\kappa_i}{\kappa_r}}\sinh\left(\frac{2\,\pi\,\kappa_i}{\kappa_r}\right) \approx\kappa_i,
	\end{split}\label{eq:evalacousticlagrangian}
\end{align}
where the approximate equivalence to $\kappa_i$ holds when $\kappa_r\approx1$ and $\kappa_i\ll1$. In eqn.\,\eqref{eq:evalacousticlagrangian}, the maximum is properly taken for $\xi\geq\pi/\kappa_r$ since we use a centered spatial average. From Fig.~\ref{fig:f01_system_scales}, it is evident that this approximation holds if the operating frequency for the acoustics is less than $25\,$GHz. It also shows that the magnitude of the Lagrangian gradient average $|\langle\partial_{\xi}\mathcal{L}\rangle_{\xi,\tau}|$ is equivalent to the amount of acoustic wave attenuation over a single wave period, $\kappa_i$. This is also clear from Fig.\,\ref{fig:f01_system_scales}.

Moreover, the exact expression in eqn.\,\eqref{eq:evalacousticlagrangian} that comes from substitution and simplification of eqn.\,\eqref{eqn:classicacousticlagrangian} also satisfies eqn.\,\eqref{eq:zero_lagrangian_identity} to good approximation all the way up to the $40\,$GHz limit for which the continuity assumption begins to break down. This can be seen in Fig.~\ref{fig:f01_system_scales}.

Finally, we consider those terms at second order in $q_{\lambda}$ and second order in combinations of $S^{-1}$ and $q_p$, leading to
%
\begin{align}\label{eq:burgers_transient_streaming_eq}
	D_tu=\mu\,\partial_x^2\,u + \eta_{\textrm{m}}^{-1}f_\text{R}(x),
\end{align}
which is a forced, viscous Burgers equation. We have dropped the streaming $(s)$ superscript indicating streaming flow at slow and relatively large scales, and we have dropped the $x$ coordinate subscript that identifies the flow as one dimensional. We have also used the nondimensional viscosity $\mu=q_{\lambda}/\RES$, itself written in terms of the streaming Reynolds number, $\RES=\rho_0\,x_s\,U_s/\mu_\text{l}$. In eqn.\,\eqref{eq:burgers_transient_streaming_eq}, we have defined
%
\begin{align}\label{eq:streaming_conversion_efficiency}
	\eta_{\textrm{m}}=(q_p\,S)^{-2}
\end{align}
as an absolute maximum streaming efficiency. It is defined in terms of the energy converted from the acoustic wave to the resulting streaming flow along the acoustic wave a distance $x$ from the source such that $\eta(x)=(u(x)/U_a)^2$. The value $\eta_{\textrm{m}}=\max_x\eta(x)$ serves as the maximum streaming efficiency possible over the entire one-dimensional acoustic streaming field. The placement of $\eta_{\textrm{m}}$ in eqn.\,\eqref{eq:burgers_transient_streaming_eq} as a coefficient on the Reynolds stress  $f_\text{R} = -\langle u^{(a)}\partial_xu^{(a)}\rangle_{\xi,\tau}$ underscores the role the Reynolds stress plays in transducing the acoustic field to the acoustic streaming flow.

\subsubsection{Steady Riccati streaming}
            
If the acoustic streaming is steady, $D_tu$ reduces to $\tfrac{1}{2}\partial_xu^2$. Then integrating eqn.\,\eqref{eq:burgers_transient_streaming_eq} over $x$ after substituting this expression produces
%
\begin{align}\label{eq:general_riccati_streaming}
\mu\,(\partial_x\,u-\partial_x\,u|_{x=0})-\tfrac{1}{2}\,u^2 = \eta_{\textrm{m}}^{-1}\int^x_0\langle u^{(a)}\partial_xu^{(a)}\rangle_{\xi,\tau}\,dx.
\end{align}
Since we have a homogeneous condition at the origin, $x=0$, $u^2|_{x=0}=0$. This equation has the form of a \citet{riccati_animadversiones_1724} equation. It also suggests that the streaming Reynolds number does not solely determine the character of the streaming flow; the nonlinearity is also important. In fact, as $q_{\lambda}\rightarrow0$, the nonlinearity plays a dominant role over viscosity in the axial acoustic streaming flow profile. If the acoustic wave is of the type given in eqn.\,\eqref{eq:stationary_acoustic_wave}, then the steady equation simplifies to
%
\begin{align}\label{eq:lin_nondiff_riccati_streaming}
	\partial_x\,u+c_s\,u^2 = c_f\,(\exp(-2\,\overline{\alpha}\,x)-1)+\partial_x\,u|_{x=0},
\end{align}
where $c_s = -(2\,\mu)^{-1}$ and $c_f=-c_s/2\,\eta_{\textrm{m}}$. A solution method exists for eqn.\,\eqref{eq:lin_nondiff_riccati_streaming} that involves transforming it into a second-order linear equation~\citep[p.\,23]{ince_ordinary_1956}. The result is
%
\begin{subequations}\label{eq:viscous_riccati_solution}
	\begin{align}
		u_{\mbox{\scriptsize visc}}&=\frac{\partial_x\,\phi}{c_s\,\phi},\\
		\phi&=I_{\beta}(h)+c_{\phi}I_{-\beta}(h),\\
		c_{\phi}&=-\frac{I_{\beta+1}(h_0)+I_{\beta-1}(h_0)}{I_{-(\beta+1)}(h_0)+I_{-(\beta-1)}(h_0)},\\
		h&=h_0\exp(-\overline{\alpha}\,x),
	\end{align}
\end{subequations}
where $I(\,\cdot\,)$ denotes the modified Bessel function of the first kind, \mbox{$\beta=h_0\sqrt{(\partial_x\,u_x|_{x=0}-c_f)/c_f}$}, and $h_0=\sqrt{(c_s\,c_f)/\overline{\alpha}^2}$. Here $\partial_x\,u|_{x=0}$ is unique and corresponds to the value that causes the solution to satisfy the far boundary condition. An exact determination of this value is not possible due to its placement within the Bessel function terms, so an iterative bisection method is necessary.
                
The steady, inviscid solution is obtained by inspection of eqn.\,\eqref{eq:lin_nondiff_riccati_streaming}: \mbox{$ u_{\mbox{\scriptsize invisc}} = \sqrt{\tfrac{1}{2\,\eta_{\textrm{m}}}(1-\exp(-2\,\overline{\alpha}\,x))}$}. Its dimensional form is
%
\begin{align}\label{eq:dimensional_inviscid_burgers_solution}
	\widetilde{u}_{\mbox{\scriptsize invisc}}=U_a\sqrt{\tfrac{1}{2}(1-\exp(-2\,\alpha\,\widetilde{x})},
\end{align}
where $\alpha=\kappa_i\,k$ is the ``true'' absorption coefficient. \citet{mitome_effects_1995} obtained a similar result in their treatment of the Rudenko and Soluyan expression. We demonstrate the broader implications of eqn.\,\eqref{eq:dimensional_inviscid_burgers_solution} in the following section.

\subsubsection{Inviscid near-source approximation}
            
Expanding in a Taylor series, we write eqn.\,\eqref{eq:dimensional_inviscid_burgers_solution} as 
\begin{equation}\label{eqn:inviscnearsource}
\widetilde{u}_{\mbox{\scriptsize invisc}}=U_a(\tfrac{1}{2}\sum_{n=1}^{\infty}\tfrac{(2\,\alpha\,\widetilde{x})^n}{n!})^{1/2}. 
\end{equation}
If we choose a point  $\widetilde{x}=\widetilde{x}_{\textrm{ns}}$ sufficiently close to the source, such that $\widetilde{x}_{\textrm{ns}}\ll1/2\,\alpha$, then we may approximate eqn.\,\eqref{eq:dimensional_inviscid_burgers_solution} as
%
\begin{align}\label{eq:near_source_approximation}
	\widetilde{u}_{\textrm{ns}} = U_a\sqrt{\alpha\,\widetilde{x}_{\textrm{ns}}},
\end{align}
describing the fluid velocity within the vicinity of the acoustic source responsible for the streaming flow. 

This result is similar to an expression produced in \citet{moudjed_scaling_2014}, where an empirical assumption was used to balance the nonlinear inertial terms with the acoustic forcing to produce the expression via a scaling argument. It is also a departure from classic theory. \cite{nyborg_acoustic_1965} applied the slow streaming assumption to the one-dimensional Eckart bulk streaming system and found that the fluid velocity $\widetilde{u}\propto\alpha$ near the acoustic source, not $\widetilde{u}\propto\sqrt{\alpha}$ as we and  \citet{moudjed_scaling_2014} found separately. In Sec.\,\ref{sec:results}, we examine past experimental results and confirm that the streaming flow dependence upon the acoustic attenuation $\alpha$ is indeed $\widetilde{u}\propto\sqrt{\alpha}$.

\subsubsection{A limit on the streaming conversion efficiency}

The maximum possible value the acoustic streaming-driven fluid flow velocity may achieve will be less than the maximum predicted flow velocity from the inviscid solution. By finding the maximum possible value of eqn.\,\eqref{eq:dimensional_inviscid_burgers_solution} over all $x$, one obtains the maximum
%
\begin{align}\label{eq:bulk_streaming_speed_limit}
	  \max_{\forall\,\sigma}\widetilde{u}<\max_{\forall\,\sigma}\widetilde{u}_\text{invisc}=U_a/\sqrt{2},
\end{align}
Moreover, by dividing eqn.\,\eqref{eq:bulk_streaming_speed_limit} by $U_a$, squaring both sides, and using eqn.\,\eqref{eq:streaming_conversion_efficiency}, we find
%
\begin{align}
	\max_{\forall\,\sigma}\left(\frac{|\widetilde{u}^{(s)}|}{U_a}\right)^2=\max_{\forall\,\sigma}\eta_{\textrm{m}}=\frac{1}{2},
	\label{eqn:maxefficiency}
\end{align}
so that the maximum efficiency possible in one-dimensional acoustic streaming is 50\%.

These limits are \emph{entirely independent of constitutive parameters}. Put another way, these are not only the maximum possible values for a given configuration, they are the maximum values for all possible configurations. That stated, the results are limited to the one-dimensional acoustic streaming configuration. It omits any consideration of multi-dimensional effects; including advection of momentum flux off-axis; acoustic beam divergence, diffraction, refraction, or focusing. It also omits and consideration of thermal effects, with the isothermal assumption. Despite these limitations, we show later that the limits do appear to apply to a broad class of acoustic streaming results. All of the constraints to this simple pair of results described above---with the exception of acoustic focusing---should produce maxima that are \emph{less} than the inviscid, isothermal, and one-dimensional values provided in eqns.\,\eqref{eq:bulk_streaming_speed_limit} and~\eqref{eqn:maxefficiency}. This means that the maximum efficiency possible for any acoustic streaming, barring acoustic focusing, is $50\,\%$. As before, there is an analogy to ordinary differential equation solutions as described in the Supplementary Information.

    \section{Results}
    \label{sec:results}
    
        When solving the Burgers partial differential eqn.\,\eqref{eq:burgers_transient_streaming_eq}, we utilize the suite of components provided by the FEniCS Project~\citep{alnaes_unified_2009,alnaes_fenics_2015,alnaes_unified_2014,kirby_algorithm_2004,kirby_compiler_2006,logg_automated_2012,logg_dolfin_2010,olgaard_optimizations_2010}. When evaluating the viscous steady Riccati solution eqn.\,\eqref{eq:viscous_riccati_solution} in the most acutely ill-conditioned cases---where both computational efficiency and arbitrary precision operation are necessary---we have employed the Advanpix Multiprecision Computing \textsc{Matlab} Toolbox~\citep{holoborodko_multiprecision_2020}.

\subsection{The maximum conversion efficiency of acoustic streaming}

\begin{figure}
	\begin{center}
		\includegraphics[width=\columnwidth]{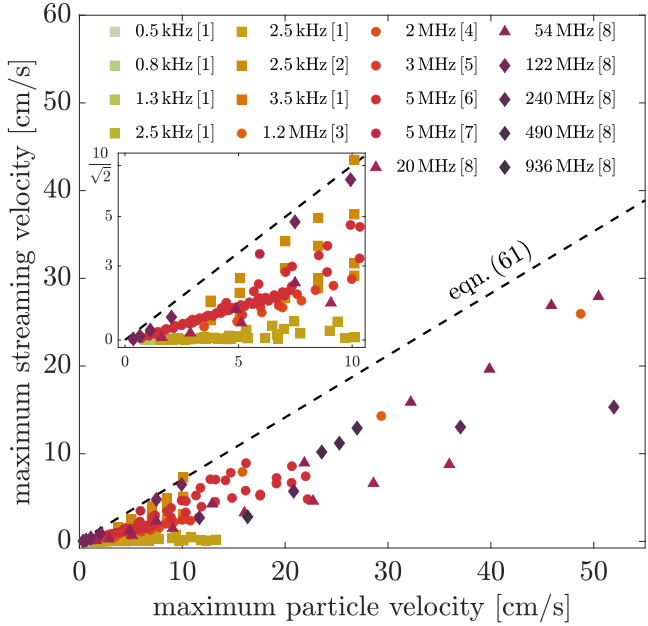}
		\caption{Bulk streaming flows bounded by the maximum streaming law, eqn.\,\eqref{eq:bulk_streaming_speed_limit}. The survey includes eighteen data sets spanning eight separate studies. Data marker types are essentially log scale in frequency: ($\square$) $f<1\,$MHz, ($\ocircle$) $f\in[1,10)\,$MHz, ($\triangle$) $f\in[10,100)\,$MHz, and ($\lozenge$) $f\in[100,1000)\,$MHz. The data are taken from [1]\,\citet{zhang_acoustic_2020}, [2]\,\citet{zhang_acoustic_2019},  [3]\,\citet{makarov_acoustic_1989}, [4]\,\citet{moudjed_scaling_2014}, [5]\,\citet{frenkel_preliminary_2001}, [6]\,\citet{mitome_mechanism_1998}, [7]\,\citet{kamakura_time_1996}, and [8]\,\citet{dentry_frequency_2014}.}
		\label{fig:f02_streaming_speed_limit}
	\end{center}
\end{figure}

We now assess the maximum achievable streaming result reported in eqn.\,\eqref{eq:bulk_streaming_speed_limit} in a comparison to experimental data reported in the literature. The selected studies report the result of bulk acoustic streaming flows that are approximately laterally unbounded and driven by plane acoustic transducers. Studies were also excluded if they failed to provide enough information to estimate both the acoustic source's particle velocity and the maximum streaming velocity. Eight separate studies~\citep{zhang_acoustic_2019,zhang_acoustic_2020,makarov_acoustic_1989,moudjed_scaling_2014,frenkel_preliminary_2001,kamakura_time_1996,mitome_mechanism_1998,dentry_frequency_2014} were chosen. They include fifteen different operating frequencies, from audible at $500\,$Hz to nearly $1\,$GHz. A plot of the maximum streaming velocity output versus maximum acoustic particle velocity as the input is provided in Fig.\,\ref{fig:f02_streaming_speed_limit} for these eight studies. An inset is provided to focus upon the more detailed results at 0--10~cm/s acoustic particle velocity. The data from these studies support the validity of the streaming law, with the exception of one data point for one study. One set of data in \citet{zhang_acoustic_2019} violates the law, though it does so in the mean value; the error bars from that study encompass the limit. Although boundary layer streaming has been excluded, it should be noted that data from such systems are expected to uniformly satisfy the law since they are characterized by the slow streaming condition.

\begin{figure}
	\begin{center}
		\includegraphics[width=\columnwidth]{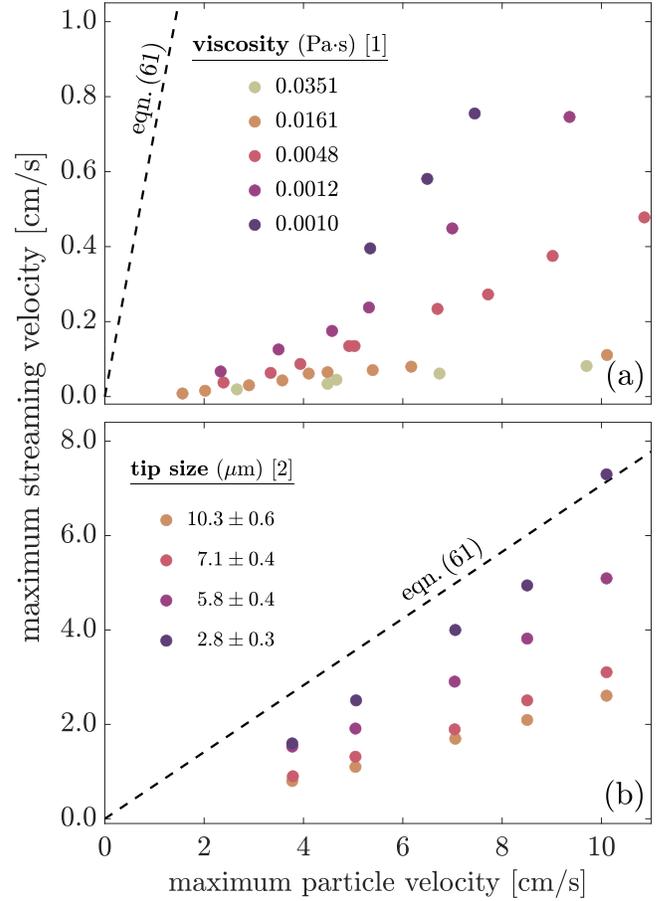}
		\caption{Data representing studies done on bulk acoustic streaming generated by vibration of a sharp tip structure at $2.5\,$kHz taken from [1]\,\citet{zhang_acoustic_2020} and [2]\,\citet{zhang_acoustic_2019}. These are special cases of the broader survey shown in Fig.\,\ref{fig:f02_streaming_speed_limit}. In subplot (a), the inverse relationship between viscosity and maximum streaming velocity aligns with the inviscid assumption used in the derivation of the streaming law. In subplot (b), tip size is twice the radius of tip curvature. As tip size is decreased, the magnitude and trend of the data approach the streaming law. This suggests that the one-dimensional model is congruent with an upper bound defined by a ``one-dimensional tip.''}
		\label{fig:f03_tip_size_and_viscosity}
	\end{center}
\end{figure}

We next specifically consider bulk acoustic streaming generated by the vibration of sharp-tipped structures with a radii of curvature (RoC) of $\leq5\,\upmu$m at $2.5\,$kHz in Fig.\,\ref{fig:f03_tip_size_and_viscosity}. These results are taken from recent, rigorously undertaken and documented studies of \citet{zhang_acoustic_2019,zhang_acoustic_2020}. Fig.\,\href{fig:f03_tip_size_and_viscosity}{3(a)} employs results from \citet{zhang_acoustic_2020} to demonstrate the effect of varying viscosity with all else (including RoC) held constant. The results indicate that decreasing viscosity in these systems tends to increase the maximum streaming velocity, bringing the data into closer agreement with eqn.\,\eqref{eq:bulk_streaming_speed_limit}, and helping to validate the approximations used in the derivation of the result and in claiming that the inviscid streaming flow would represent the upper bound of all streaming flows. In Fig.\,\href{fig:f03_tip_size_and_viscosity}{3(b)}, we have plotted results from the earlier work of \citet{zhang_acoustic_2019} that characterize the effect of tip size, where they define the tip size as $2\,$RoC. The data show that as the tip size is decreased, the profile approaches the streaming law, both in terms of magnitude and in terms of trend. This suggests congruence of the one-dimensional model with a physical ``one-dimensional tip'' system. The single data point that weakly violates the streaming law is the same that was previously noted in Fig.\,\ref{fig:f02_streaming_speed_limit}.

\subsection{Shared physics of bulk streaming systems}

\begin{figure}
\begin{center}
\includegraphics[width=\columnwidth]{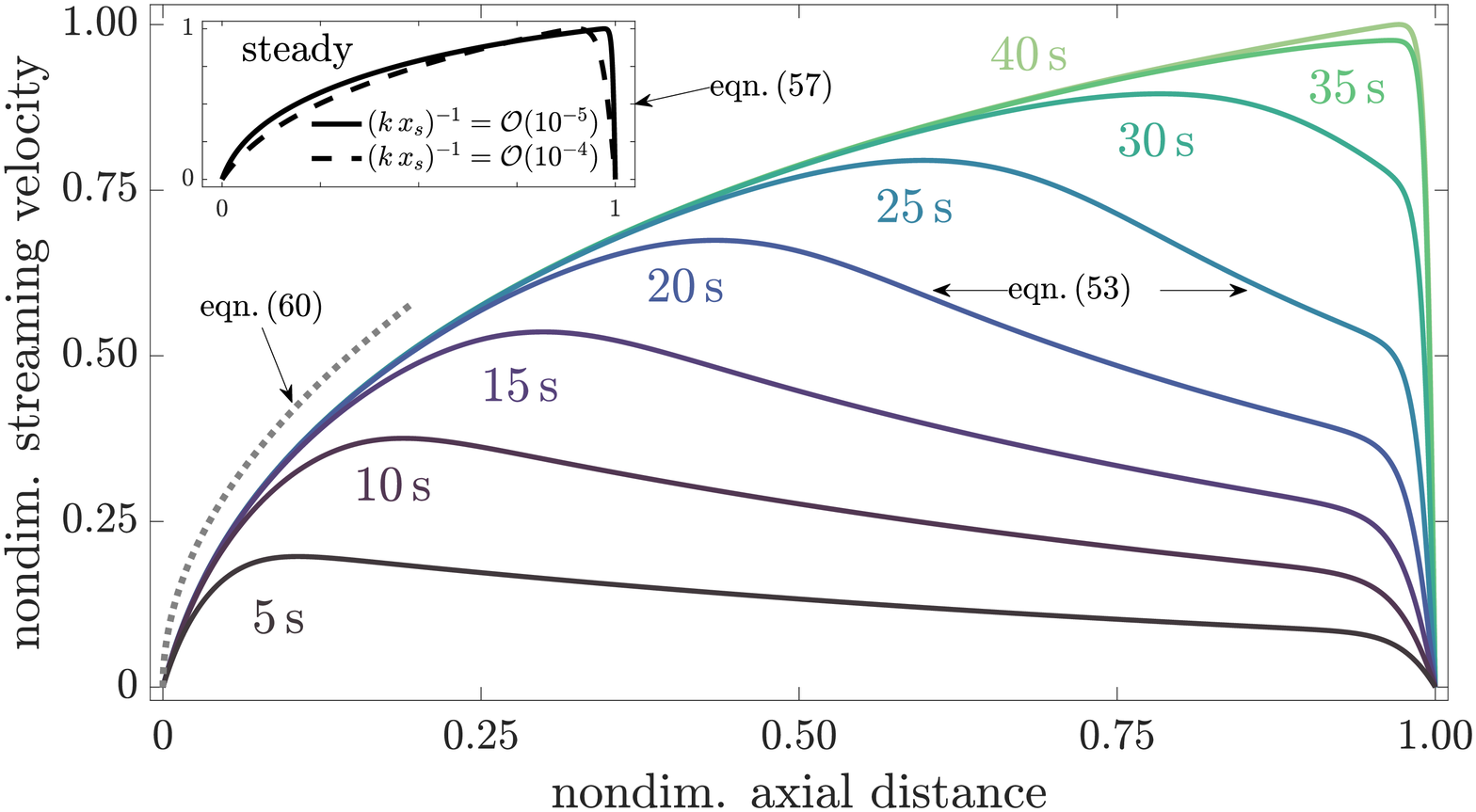}
\caption{Fast axial Eckart streaming with $\mu^{-1}=7720$ and $\eta_{\textrm{m}}=0.3$. The acoustic forcing is provided by a linear, non-diffracting wave with the attenuation coefficient $\alpha=0.172$. In the inset, steady streaming for (solid) as system with the noted parameters is compared with (dashed) a system where domain length has been shortened (with parameters otherwise equivalent between the two systems). The domain length of the former is such that $q_{\lambda}=(k\,x_s)^{-1}=\mathcal{O}\left[10^{-5}\right]$, whereas the domain length of the latter is shorter, such that $q_{\lambda}=(k\,x_s)^{-1}=\mathcal{O}\left[10^{-4}\right]$. The change from one condition to the other alters the importance of each term on the left hand side of eqn.\,\eqref{eq:general_riccati_streaming}. In the main plot, the profile shape in the vicinity of the source agrees with the near-source approximation (dotted) for the longer domain. This is due to the greater degree of nonlinear profile development within the longer domain.}
\label{fig:f05_eckart_streaming}
\end{center}
\end{figure}

\citet{eckart_vortices_1948} streaming is the net bulk flow generated by an acoustic source within a bounded fluid medium when the far boundary is a perfect acoustic absorber (Fig.\,\href{fig:f04_streaming_types}{4(a)}). Equation~\eqref{eq:burgers_transient_streaming_eq} provides a new model for the transient layering behavior that is empirically observed in such systems~\citep{kamakura_time_1996,moudjed_near-field_2015}. The viscous flow solution in eqns.\,\eqref{eq:viscous_riccati_solution} defines a ``shark fin'' (\ie, rounded sawtooth) spatial profile that is characteristic of steady axial Eckart flow. In the limit of the inviscid approximation, the Eckart system ``loses'' the distal boundary condition and the result can be interpreted as the bulk flow solution. We will show how this inviscid solution is related to Stuart-\citet{lighthill_acoustic_1978} ``jet'' streaming (Fig.\,\href{fig:f04_streaming_types}{4(b)}). All of the noted features from these streaming phenomena are shown in Fig.\,\ref{fig:f05_eckart_streaming}.
\begin{figure}
\begin{center}
\includegraphics[width=0.75\columnwidth]{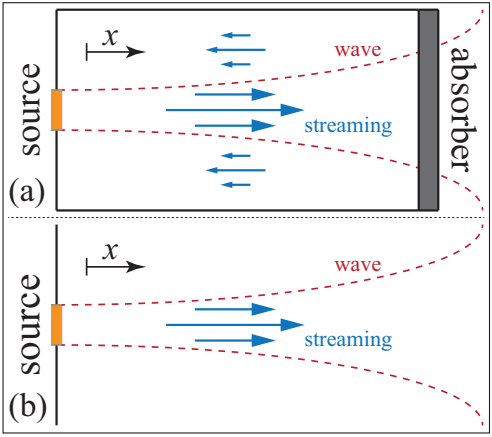}
\caption{Simplified diagram depicting bulk acoustic streaming types. (a) Eckart streaming and (b) Stuart-Lighthill ``jet'' streaming. In either scenario, the near-source behavior is that of a McIntyre sink~\citep{lighthill_acoustic_1978}. In (a), the continuation of the acoustic wave beyond the boundary is representative of the fact that one can replace the perfect absorber with an acoustically transparent membrane (\eg polyester (Mylar$^\text{\textregistered}$) membrane \cite{lee_calibration_2010}) to achieve a similar effect, as illustrated in \citet{nyborg_acoustic_1965}.}
\label{fig:f04_streaming_types}
\end{center}
\end{figure}
                
Lighthill's seminal acoustic streaming study~\citep{lighthill_acoustic_1978} is one of the most well-known and oft-cited works on the topic. It addresses, among other subjects, the description of bulk turbulent jet streaming within an unbounded medium (Fig.\,\href{fig:f04_streaming_types}{4(b)}). The jet streaming is generated from high-frequency ($\geq1\,$MHz) acoustic forcing that Lighthill treated as a point source in light of the relatively large associated attenuation coefficient. In the modern applied acoustofluidics setting, flow dynamics adjacent the acoustic source and well within the attenuation length scales are important, and so the source must be more carefully treated. The point-source treatment invokes a singularity as $x\rightarrow 0^+$ that has no observable physical analog. Experiments by \citet{dentry_frequency_2014} reveal the presence of zero fluid velocity at the source with algebraic growth of the velocity near the source that is mediated in the far field with a long-tailed decay, producing a well-defined maximum.
            
In a recent study, \citet{dentry_frequency_2014} attempted to rectify the weaknesses of the point-source approximation by modifying Lighthill's model to account for both a finite source area and for laminar streaming. Laminarity is evident in Dentry's study by direct observation. While the study provides useful insights toward achieving its broader objectives, it is limited by a number of errors that stem from two important oversights:
%
\begin{enumerate*}[label=(\roman*)]
	\item\label{it:error_source_1} an incorrectly stated equation that was later discussed (though not completely addressed) in a published erratum~\citep{dentry_erratum_2016}, and 
	\item\label{it:error_source_2} the determination of streaming flow with particle image velocimetry (PIV) by utilizing $a\approx5\,\upmu$m diameter particles with very high frequency (small wavelength) acoustic sources.
\end{enumerate*}
Errors associated with item \ref{it:error_source_1} culminate in a correction factor of $\approx2.65$ multiplying the reported acoustic power (with respect to particle velocity or displacement, the correction is quadratic rather than linear). Additionally, the errors affect the model's accuracy and raise questions as to whether the data is correctly represented in the study. Error source \ref{it:error_source_2} is important because, for $a\gtrsim\lambda$, effects of the acoustic radiation force on the particle become significant and the particle does not reliably follow the streaming field. This invalidates the data reported in \citep{dentry_frequency_2014} for $f\geq240\,$MHz, which equates to a third of the experimental data in that work.
            
As a basis for comparison to our analysis, we use the general form of the Lighthill and Dentry models for bulk jet streaming into an unbounded medium:
%
\begin{align}\label{eq:lighthill_dentry}
	\widetilde{u}_d(\widetilde{r},\widetilde{x}) &= \sqrt{\frac{2\,P}{\pi\,\rho\,c\,S(\widetilde{x})^2}}\sqrt{1-\exp(-2\,\alpha\,\widetilde{x})}\exp\biggl[-\left(\frac{\widetilde{r}}{S(\widetilde{x})}\right)^2\biggr],
\end{align}
where $\widetilde{r}$ denotes the radial coordinate. For simplicity we have assumed an ideal (\ie, uniform acoustic intensity) circular thickness-mode transducer. The length $S$ is obtained as the numerical solution to a stiff ordinary differential equation. It accounts for the spatial rate of lateral losses by normalizing a transverse Gaussian decay. The length $S$ also enforces far field axial velocity decay via multiplicative attenuation.

If we ``eliminate'' the lateral dimensionality by considering only the jet axis ($\widetilde{r}=0$) and fixing $S=2\sqrt{A/\pi}$, then the jet-streaming model of eqn.\,\eqref{eq:lighthill_dentry} simplifies to become identical to the inviscid Riccati solution of eqn.\,\eqref{eq:dimensional_inviscid_burgers_solution}. In fact, Lighthill's semi-empirical derivation initially produces a one-dimensional model of exactly this form. He subsequently extends the model to include lateral dimensionality by assuming a Gaussian decay of the streaming velocity away from the jet axis. 

The fundamental difference in our approach is that we have derived eqn.\,\eqref{eq:dimensional_inviscid_burgers_solution} directly from the Navier-Stokes equations. Its form originates in the nonlinear terms of eqn.\,\eqref{eq:burgers_transient_streaming_eq}, and these are regarded as dominant in our partitioning analysis as a direct consequence of the large-amplitude streaming assumption.

We avoid the computational demand of Lighthill's approach by  directly obtaining a highly-simplified algebraic closed-form model for the axial streaming profile from the steady-state Riccati physics. We achieve this by first deriving a scaling relation between the power and the axial location of the streaming maximum (with Buckingham's $\Pi$ theorem~\cite{buckingham_physically_1914}). This is used in conjunction with an attenuation prefactor and boundary conditions to account for advection of momentum away from the jet axis. The details are included in the Supplementary Information. The analysis is carried out with the valid PIV regime data ($f\leq122\,$MHz) from \citet{dentry_frequency_2014}. Where necessary, the data and models from that study are corrected according to the associated erratum~\citep{dentry_erratum_2016}. The result of these efforts is the inviscid model of eqn.\,\eqref{eq:dimensional_inviscid_burgers_solution} modified by an attenuation prefactor:
%
\begin{align}\label{eq:empirical_jet}
	\widetilde{u}_j=B(\widetilde{x})\,\widetilde{u}_{\mbox{\scriptsize invisc}},
\end{align}
where
%
\begin{align}
	B(\widetilde{x}) = \frac{1-\exp(2\,\alpha\,x_s)+\alpha\,x_s}{1-\exp(2\,\alpha\,x_s)+\alpha\,(x_s-\widetilde{x})},
\end{align}
and
%
\begin{align}\label{eq:buck_pi_estimate}
	x_s = \frac{34\,P^{0.1}}{\rho_0^{0.1}\,\alpha^{0.5}\,f^{0.3}},
\end{align}
is the Buckingham-$\Pi$ estimate of the axial distance to maximum streaming velocity. The accuracy of eqn.\,\eqref{eq:buck_pi_estimate} is illustrated in Supplementary Information Fig.\,2. The boundary conditions are chosen so that $\widetilde{u}_j$ attains its maximum at $x_s$ and the near-source, inertia-dominant regime is asymptotically equivalent to eqn.\,\eqref{eq:near_source_approximation}. In other words, the model is defined by steady Riccati physics.

\begin{figure*}
\begin{center}
\includegraphics[width=\textwidth]{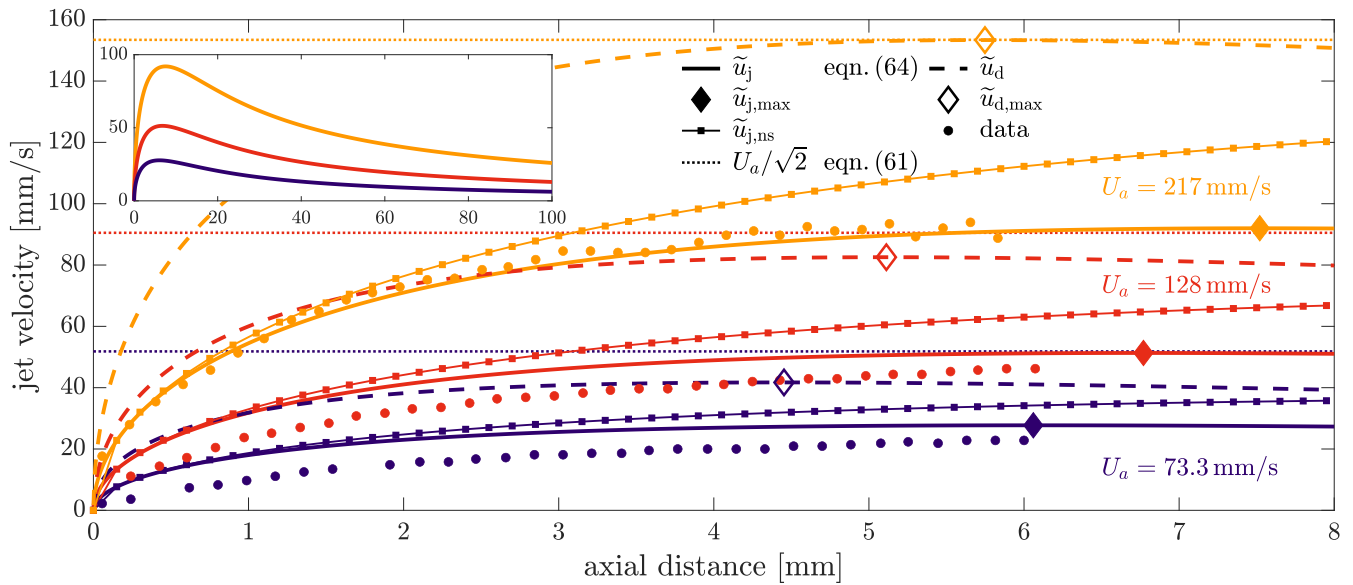}
\caption{Stuart-Lighthill jet streaming axial profiles at three different acoustic forcing magnitudes. Data and reference (laminar) models are taken from \citet{dentry_frequency_2014} with corrections applied according to the associated erratum \citep{dentry_erratum_2016}. The attenuation prefactor used with $\widetilde{u}_j$ is derived to satisfy two boundary conditions. Near the source, the profile agrees asymptotically with eqn.\,\eqref{eq:near_source_approximation}. Away from the source, the profile maximum is enforced to occur at $x_s$, where $x_s$ is estimated from the Buckingham $\Pi$ scaling relation of eqn.\,\eqref{eq:buck_pi_estimate}. Thus, the curvature, the maximum amplitude, and the location of the maximum are systematically enforced from eqn.\,\eqref{eq:dimensional_inviscid_burgers_solution}. A high-level perspective of the efficacy of this approach is provided in Fig.\,\ref{fig:f07_jet_maximum}.}
\label{fig:f06_jet_profile}
\end{center}
\end{figure*}

The leading portion of the axial jet profile (from source to maximum) is examined in Fig.\,\ref{fig:f06_jet_profile}. This portion of the flow is a mechanistic analog for steady Eckart streaming through eqn.\,\eqref{eq:dimensional_inviscid_burgers_solution}. The models and data are shown for fixed frequency at three different source velocities. The near-source jet velocity, $\widetilde{u}_{j,\mbox{\scriptsize ns}}=B(x)\,\widetilde{u}_{\mbox{\scriptsize ns}}$, is also shown. Aside from the obvious improvement in amplitude correspondence, eqn.\,\eqref{eq:empirical_jet} enforces a downstream shift in the maximum streaming velocity distance estimate. Since the data are monotonic with an apparent nonzero slope, the shift represents an increase in accuracy. Within the vicinity of the source, the profile curvature is evidently better explained by the closed-form model. This region is strongly correlated to the near-source approximation, which itself is from the inviscid Riccati solution. At smaller amplitudes, the range of validity of the near-source approximation extends roughly to $x_s$.
            
\begin{figure*}
\begin{center}
	\includegraphics[width=\textwidth]{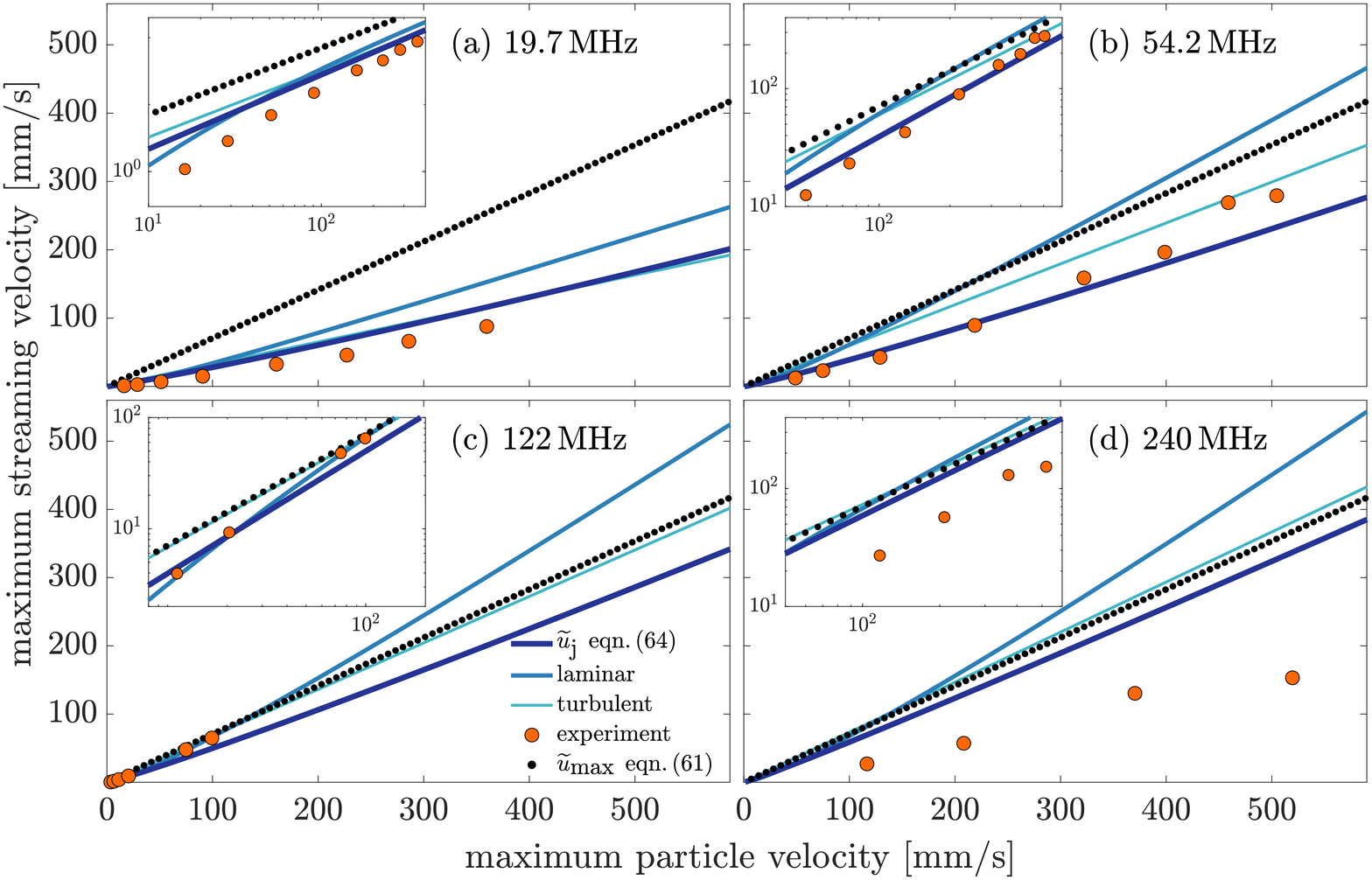}
	\caption{Maximum velocity profiles for steady Stuart-Lighthill jet streaming at differing acoustic forcing frequencies. Overall, eqn.\,\eqref{eq:empirical_jet}---with attenuation properties enforced to satisfy eqns.\,\eqref{eq:near_source_approximation} and eqn.\,\eqref{eq:buck_pi_estimate}---better accounts for the observed behaviors. Its algebraic form ensures computational efficiency. The streaming law is evidently violated by the Lighthill and Dentry models, while eqn.\,\eqref{eq:empirical_jet} adheres to the law by construction. The systematically low-valued data from \citet{dentry_frequency_2014} in subplot (d) is likely spurious as a result of significant acoustic radiation forcing. This is due to the relatively small acoustic wavelength---$\lambda\lesssim6\,\upmu$m for $f\gtrsim240\,$MHz---compared to the PIV particle size, $5\,\upmu$m, used in that study.}
	\label{fig:f07_jet_maximum}
\end{center}
\end{figure*}

The streaming maxima for the experimental data and our models are plotted together as a continuous function of maximum particle velocity in Fig.\,\ref{fig:f07_jet_maximum}. Overall, the result of our closed-form model provides the most faithful characterization of the experimental observations. It does so at significantly less computational cost, since the other models involve the numerical integration of stiff, nonlinear ordinary differential equations. The systematically low-valued data in Fig.\,\href{fig:f07_jet_maximum}{7(d)} are an inaccurate accounting of the streaming flow due to the effects of direct acoustic radiation forcing on the $5\,\upmu$m PIV particles.

\subsection{Transient development of bulk acoustic streaming}

The transient flow development of the Stuart-Lighthill jet \citep{lighthill_acoustic_1978} can be approximated by using our closed form model in conjunction with the inviscid Burgers equation---eqn.\,\eqref{eq:burgers_transient_streaming_eq} with $\mu$ set to zero. We consider the leading portion of the jet profile (\ie, from the source to the axial coordinate of the velocity maximum). The transient component of the streaming field is extracted with the nondimensional transient energy density:
%
\begin{align}\label{eq:transient_streaming_energy_density}
	\langle\mathcal{E}_{\delta u}\rangle_{x}(t)=\frac{1}{2}\int^1_0\delta u(x,t)^2\,dx,
\end{align}
%
\begin{figure*}
\begin{center}
	\includegraphics[height=8 cm]{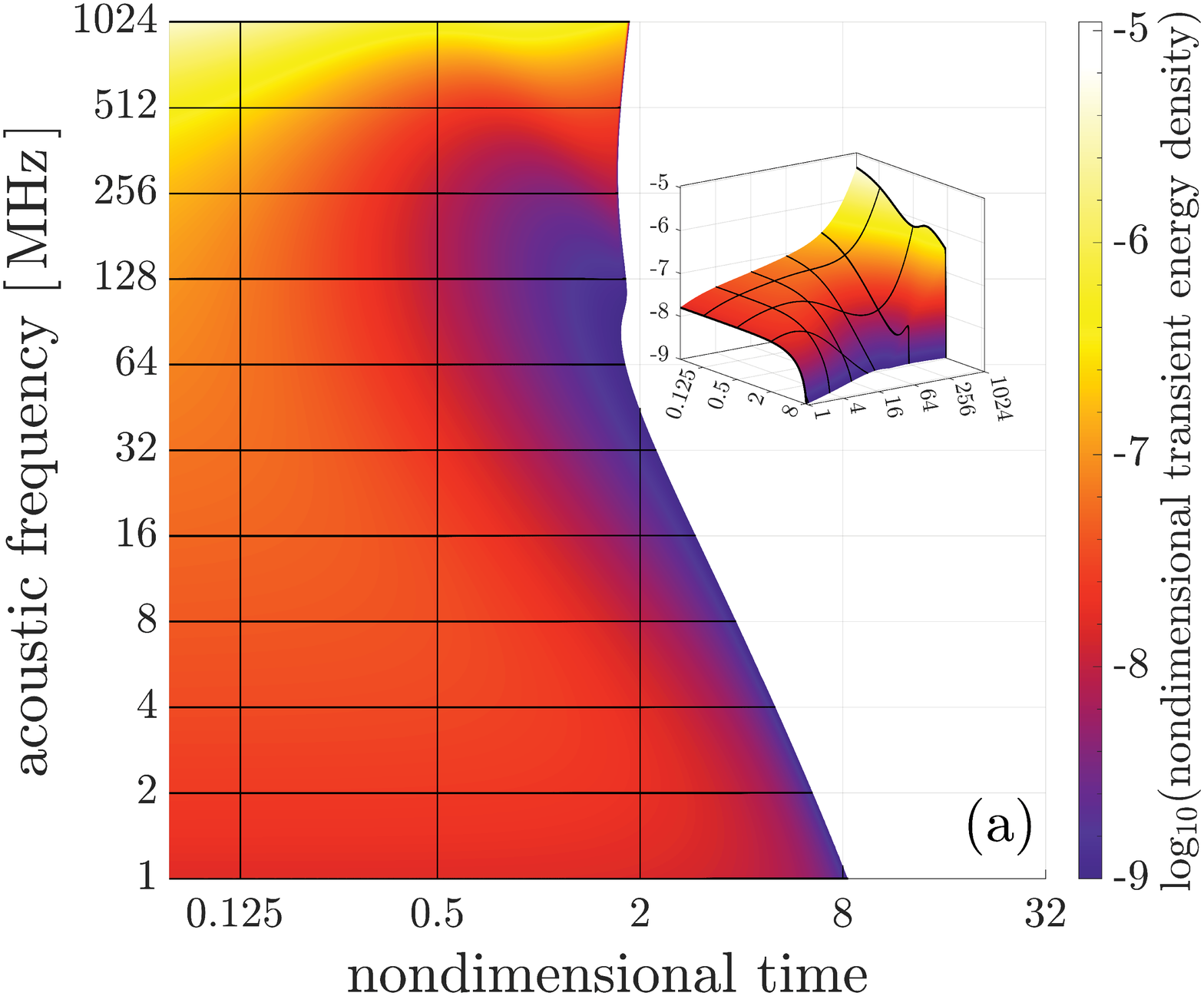}
	\includegraphics[height=8 cm]{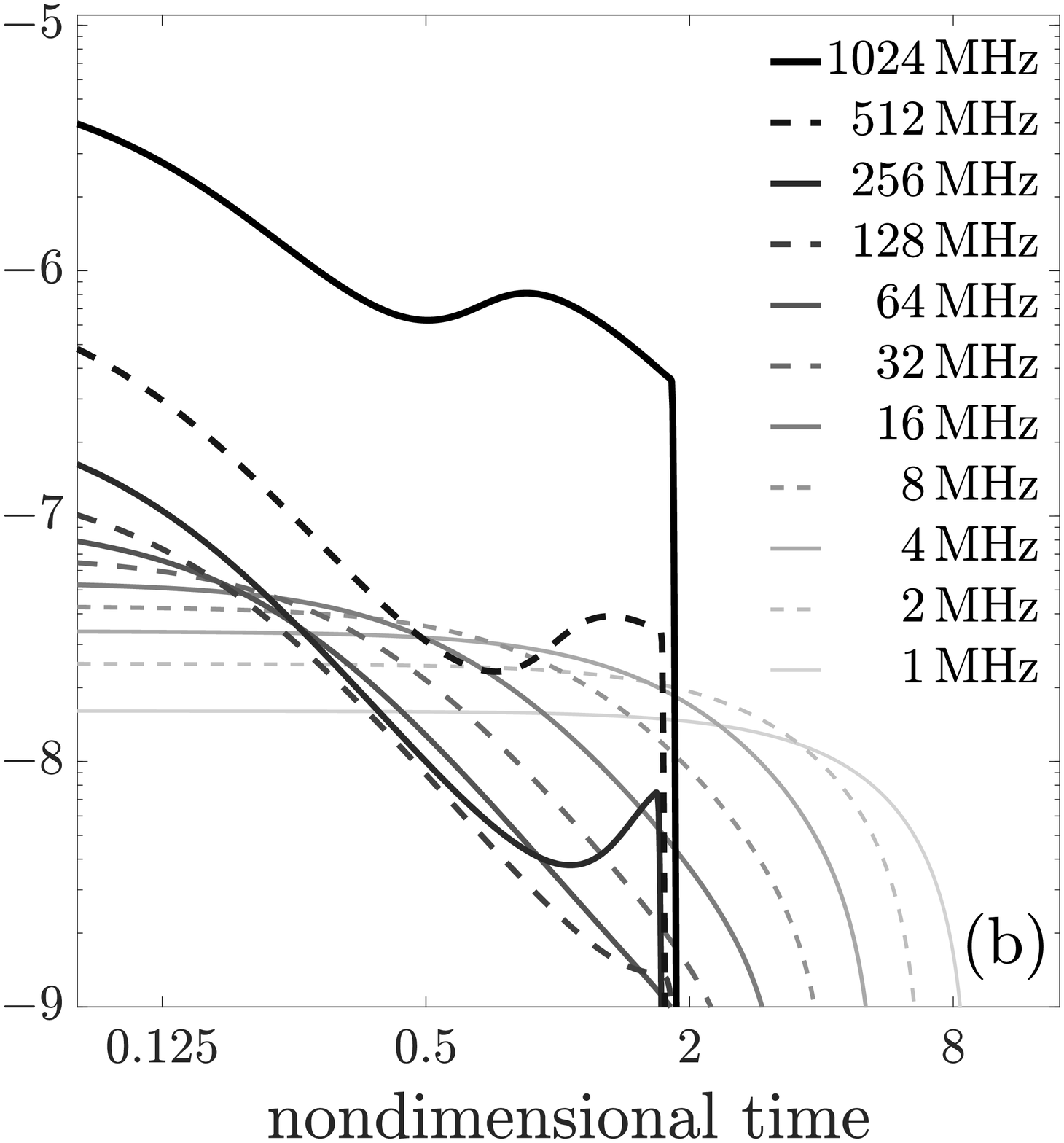}
	\caption{Transient energy density for the leading portion of the Stuart-Lighthill jet profie. The transient dynamics are extracted from the net streaming flow with definition Eq.\,\eqref{eq:transient_streaming_energy_density}. In subplot (a), time to steadiness is a non-monotonic function of frequency exhibiting trend reversals at $\sim100\,$MHz and $\sim256\,$MHz. Frequency isolines in the surface inset correspond to solid profiles in plot (b), where three regimes are observed. At low frequencies, evenly distributed transient energy rolls off quickly into steadiness. At moderate frequencies, power-law transient decay gives way to energy peak development at the terminal end and instantaneous decay to steadiness. At high frequencies, the peak shifts away from the terminal edge and defines a local maximum.}
	\label{fig:f08_transient_energy_density}
\end{center}
\end{figure*}
where the nondimensional background density is unity, $dV=dx$, and \mbox{$\delta u(x,t)=u(x,t)-u(x,t-dt)$} is the transience of the Burgers streaming field. The transient flow behaviors are brought to light in Fig.\,\ref{fig:f08_transient_energy_density}. The most striking features are found where the \emph{steadiness}, formally defined as the nondimensional transient energy density, $\langle\mathcal{E}_{\delta u}\rangle_x<10^{-9}$, is nonmonotonically changing with respect to time. These changes are especially seen as the acoustic frequency is increased beyond 100~MHz. In Fig.\,\href{fig:f08_transient_energy_density}{8(a)}, two trend reversals are observed, a minor reversal at $\sim100\,$MHz and a major reversal at $\sim256\,$MHz. At these frequency values, the time to achieve steady acoustic streaming is at a minimum nondimensional time. Above and below these values, it takes longer to achieve steady acoustic streaming. Above 256~MHz in particular, time to steadiness appears to be an increasing function of frequency, despite a fixed on-source particle velocity and decreasing vibrational amplitude. 

A corresponding trend reversal is observed in the frequency isolines of the inset surface plot. The isolines correspond to the profiles plotted in Fig.\,\href{fig:f08_transient_energy_density}{8(b)} and reveal three regimes. At frequencies below the minor reversal, transient energy is more evenly distributed and exhibits smooth rolloff into steadiness. As the frequency is increased to within the minor reversal, the transient decay is linear in log-log space (\ie, follows a power law, see $64\,$MHz and $128\,$MHz). This culminates in the formation of an energy peak near the terminal profile edge for frequencies approaching the major reversal. With peak formation, decay to steadiness evolves from ``smooth and slow'' toward ``sharp and instantaneous.'' Above $\sim256\,$MHz, the energy peak shifts away from the steady edge, leading to the development of a local maximum.

\subsection{The effect of frequency on steady bulk acoustic streaming}

\begin{figure}
	\begin{center}
		\includegraphics[width=\columnwidth]{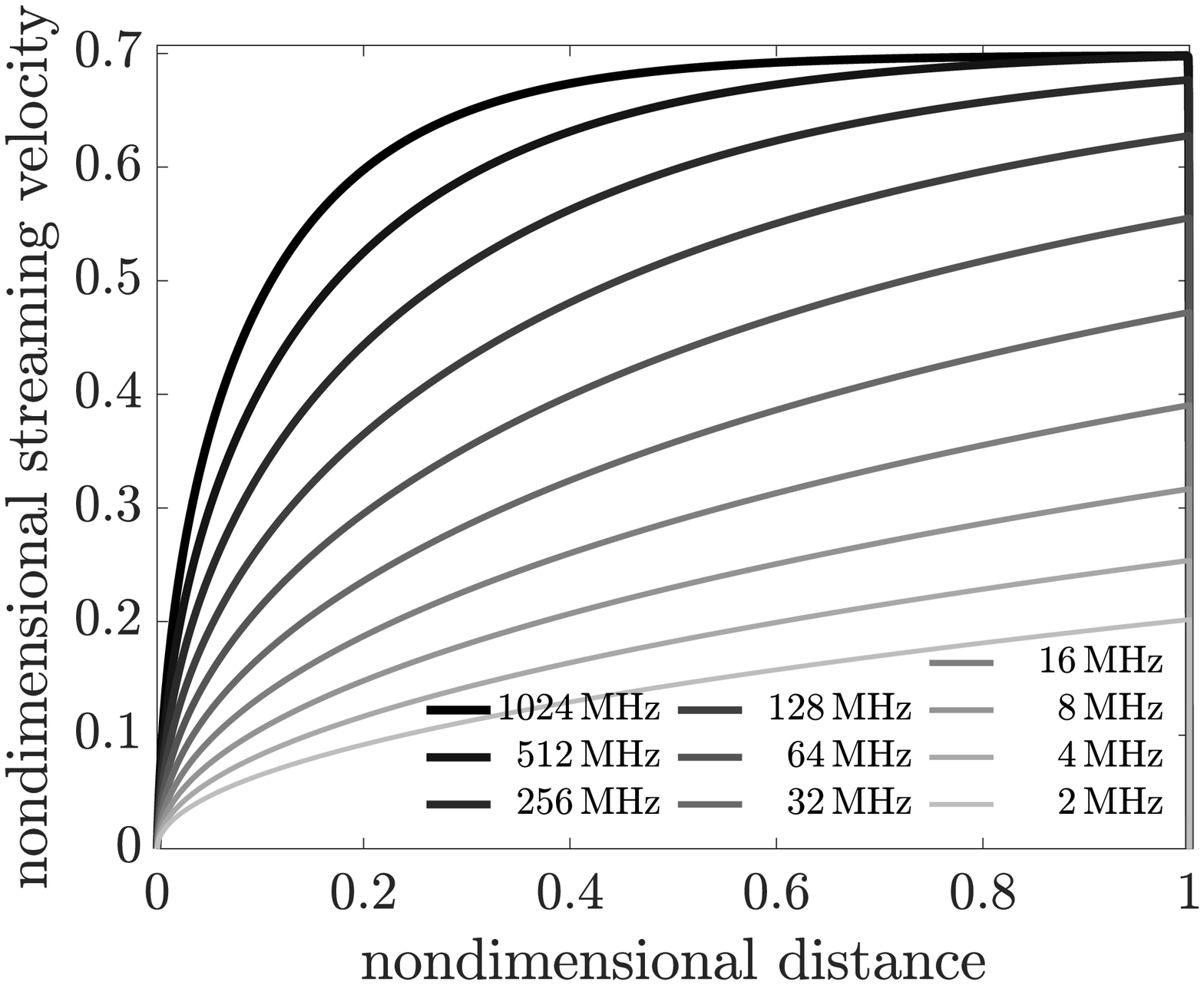}
		\caption{Steady streaming velocity profiles for a logarithmic frequency space defined over a fixed acoustic intensity (and power). Only the highest frequency attains the theoretical maximum streaming velocity in the finite domain. Though the maximum streaming velocity is itself directly proportional to the maximum vibration amplitude, the evolution of the jet velocity over the distance from the acoustic source depends on the Reynolds stress, as shown in eqn.\,\eqref{eq:dimensional_inviscid_burgers_solution}. The Reynolds stress term attains its maximum as the square root of a Gaussian function of increasing frequency, since $\alpha\propto\omega^2$.}
		\label{fig:f09_steady_profiles}
	\end{center}
\end{figure}

We finally consider the effect of the frequency on the steady streaming profile in Fig.\,\ref{fig:f09_steady_profiles}. This is achieved by using the closed-form model while maintaining a constant acoustic intensity (and power) from the source. With the assumption of uniform intensity over the emitting surface, we have $U_a=2\,\pi\,f\,\xi_p=\sqrt{\tfrac{I}{z_0}}=\sqrt{\tfrac{P/A}{z_0}}$ for a device of fixed area $A$ and with $z_0=\rho_0\,c$ as the acoustic impedance of the fluid. The profiles in the plot demonstrate that the theoretical maximum is achieved over a shorter distance as frequency is increased and power is held constant, as one would probably expect. Over a majority of the domain, the streaming velocity is approximately linearly incremented in response to a logarithmically incremented frequency grid. Since the streaming speed limit in eqn.\,\eqref{eq:bulk_streaming_speed_limit} is non-constitutive, the observed relationship implies that, for inviscid systems where the approximation eqn.\,\eqref{eq:dimensional_inviscid_burgers_solution} is valid, the only means for attaining the theoretical maximum streaming velocity within a finite domain is by increasing the acoustic frequency. Though the maximum streaming velocity is directly proportional to the vibrational amplitude, the exponential factor representing the Reynolds stresses attains its maximum value as the square root of a Gaussian function of increasing frequency ($\alpha\propto\omega^2$);  it is independent of vibrational amplitude.

\section{Concluding Remarks}
\label{sec:conclusions}

Over the last several decades, advancements in acoustofluidics have brought within practical reach many useful applications across disciplines. The rapid progress of these innovations has outpaced theory and left behind many unresolved questions. This shortage of understanding arises due to the intractable governing nonlinear equations of motion. Classical efforts at addressing these difficulties were introduced by Lord Rayleigh in his early investigations of acoustic streaming nearly one hundred fifty years ago.

Rayleigh's methods provided the template for the development of similar techniques over the years to the modern era by notable acousticians and fluid mechanists. The common feature in all these approaches is the assumption that the acoustic streaming flow is much smaller in magnitude than the driving acoustics. In the more than forty years since Lighthill showed that this assumption cannot be generalized to all acoustofluidic settings, little progress has been made toward a more general, systematic approach.

This study has detailed a much needed alternative to the traditional perturbative technique. Our systematic approach has flexibility and generality as its foundation. The method adopts the unconventional strategy of differentiating across the vast spatiotemporal scale disparities encountered in high-frequency acoustofluidics, including not only a separation in temporal scales but also spatial scales between the acoustics and resulting hydrodynamics. The key result of this method is a field-decoupled, finite-term expansion of the governing equations, with terms stratified by order of importance as determined by the scale disparities. This approach allows for identification and segmentation of specific flow structures, as designated by the user when assigning characteristic scales.

We have investigated the usefulness of the method by applying it to a simple one-dimensional large-amplitude bulk streaming system. Definition of the scales in this case is based on an axiomatic assumption of commensurately ordered acoustic particle velocities and streaming flow velocities. We have shown that for valid inclusion of the spatial acoustic averaging constraint, the wavelength of the acoustic forcing should be less than the attenuation length. It was also revealed that this condition is approximately satisfied by our approach to the analysis across the entire domain of continuum mechanics. The nonlinear equations of motion approximating fast axial streaming were recovered, and it was shown that, for the one-dimensional problem, the transient system is written as a viscous Burgers equation forced by the Reynolds stresses of the acoustic wave. Solutions describing the transient onset of acoustic streaming were explicitly derived for the first time. The Burgers description reduces, at steady state, to a Riccati equation. We have thoroughly characterized the Burgers-Riccati system as it relates to Eckart streaming, Stuart-Lighthill streaming, and other recently studied forms of bulk acoustic streaming. In the case of a bounded domain, it was revealed that the extent of streaming nonlinearity depends on the ratio of the wavelength to the domain length in addition to viscosity, and that this is also a factor in determining the maximum streaming velocity. Using an inviscid approximation, we derived simple expressions for the near-source inertial streaming behavior and the maximum attainable streaming velocity. For the latter result, it was shown that on a ``short'' bounded domain, the only means of asymptotically achieving the theoretical maximum streaming velocity is by increasing the frequency of the acoustic forcing. From the theoretical maximum streaming law, we recovered a universal upper bound on the energetic conversion efficiency of $50\,\%$, independent of constitutive parameters. These findings were rigorously validated by comparison to theoretical and experimental findings from a broad survey of the literature.

\begin{acknowledgments}
    The work presented here was generously supported by a SERF research grant to J.\ Friend from the W.\ M.\ Keck Foundation. He is furthermore grateful for the support of this work by the Office of Naval Research (Grant No.\ 12368098). J.\ Orosco is thankful for support provided by the University of California's Presidential Postdoctoral Fellowship Program.
\end{acknowledgments}

\bibliographystyle{apsrev4-1}
\bibliography{bib}
\end{document}